\input harvmac.tex
\def\frac#1#2{{#1\over#2}}
\def\half{\frac12}
\def\exp{{\rm exp}}

\def\l{\lambda}

\def\cQ{{\cal Q}}

\def\l{\lambda}

\def\t{\theta}

\def\hf{{1\over2}}


\lref\MKLas{Kedem, R., Klassen, T.R., McCoy B.M. and Melzer E.:
Fermionic sum representations for conformal field theory 
characters. Phys. Lett. {\bf B307}, 68-76 (1993)}

\lref\McCoy{McCoy, B.M.: The connection between
statistical mechanics and Quantum
Field Theory.
In: Bazhanov, V.V. and Burden, C.J. (eds.)
Field Theory and Statistical Mechanics. Proceedings 7-th Physics
Summer School at the Australian National University. Canberra. 
January 1994, World Scientific 1995 }

\lref\DotsFat{ Dotsenko, Vl.S. and Fateev, V.A.:
Conformal algebra and
multipoint correlation functions in 2d statistical models. Nucl. Phys.
{\bf B240}\ [{\bf FS12}], 312-348 (1984)\semi
Dotsenko, Vl.S. and  Fateev, V.A.: Four-point
correlation functions and the operator algebra in 2d conformal invariant
theories with central charge $c\le1$.
Nucl. Phys. {\bf B251}\ [{\bf FS13}] 691-734 (1985) }

\lref\Zkvadrat{Zamolodchikov, A.B. and Zamolodchikov, Al.B.:
Factorized S-matrices in two dimensions as the exact
solutions of certain relativistic quantum field theory models.
Ann. Phys. (N.Y.) {\bf 120}, 253-291 (1979) }

\lref\Fei{Feigin, B.L. and  Fuchs, D.B.  Representations
of the Virasoro algebra. In:
Faddeev, L.D., Mal'cev, A.A. (eds.) Topology.
Proceedings, Leningrad 1982.
Lect. Notes in Math.
{\bf 1060}.
Berlin, Heidelberg, New York: Springer 1984}

\lref\RKSA{Kulish, P.P., Reshetikhin, N.Yu. and  Sklyanin, E.K.:
Yang-Baxter equation and representation theory. Lett. Math.
Phys. {\bf 5}, 393-403 (1981) }

\lref\Zar{Zamolodchikov, Al.B.: Thermodynamic Bethe ansatz in
relativistic models: Scaling 3-state Potts and Lee-Yang models.
Nucl. Phys. {\bf B342}, 695-720 (1990)}

\lref\YY{Yang, C.N. and Yang, C.P.: Thermodynamics of one-dimensional
system of bosons with repulsive delta-function potential.
J. Math. Phys. {\bf 10}, 1115-1123 (1969)}

\lref\FSLS{Fendley, P., Lesage, F.  and Saleur, H.:
Solving 1d plasmas and 2d boundary problems using Jack
polynomial and functional relations. Preprint
USC-94-16, SPhT-94/107,
$\#$hepth 9409176 (1994)}

\lref\Zar{Zamolodchikov, Al.B.: Thermodynamic Bethe ansatz in
relativistic models: Scaling 3-state Potts and Lee-Yang models.
Nucl. Phys. {\bf B342}, 695-720 (1990)}

\lref\ABZ{Zamolodchikov, A.B.:
Integrable field theory from conformal field theory.
Adv. Stud. in Pure Math.
{\bf 19}, 641-674 (1989)}

\lref\GZ{Ghosal, S. and Zamolodchikov, A.B.:
Boundary S-matrix and boundary state in two-dimensional integrable
quantum field theory.
Int. J.  Mod. Phys. {\bf A9}, 3841-3885 (1994)}

\lref\FadS{Faddeev, L.D., Sklyanin, E.K. and Takhtajan, L.A.:
Quantum inverse scattering method I.
Theor. Math. Phys. {\bf 40}, 194-219 (1979) (in Russian)}

\lref\dVega{de Vega, H.J. and Destri., C.:
Unified approach to thermodynamic Bethe Ansatz and finite size corrections
for lattice models and field theories. 
Nucl. Phys. {\bf B438}, 413-454 (1995)}

\lref\jap{Sasaki, R. and Yamanaka, I.:
Virasoro algebra, vertex operators, quantum Sine-Gordon
and solvable Quantum Field theories. Adv. Stud. in Pure Math.
{\bf 16}, 271-296 (1988) }

\lref\BLZ{Bazhanov, V.V., Lukyanov, S.L. and Zamolodchikov, A.B.:
Integrable structure of conformal field theory,
quantum KdV theory and
thermodynamic Bethe ansatz.
Preprint CLNS 94/1316, RU-94-98,
\#hepth  9412229}

\lref\Kor{Bogoliubov, N.M., Izergin, A.G. and  Korepin, V.E.:
Quantum Inverse Scattering Method
and Correlation Functions.  Cambridge: University Press 1993}

\lref\MKlas{Kedem, R., Klassen, T.R., McCoy B.M. and Melzer E.:
Fermionic sum representations for conformal field theory 
characters. Phys. Lett. {\bf B307}, 68-76 (1993)}

\lref\Zamt{Zamolodchikov, Al.B.: Private communication.}

\lref\Baxter{Baxter, R.J.: Exactly Solved Models
in Statistical Mechanics. London: Academic Press 1982}

\lref\ALZ{Zamolodchikov, Al.B.:
Painlev$\acute{\rm  e}$ III and 2D polymers,
Nucl. Phys. {\bf B432}, 427-456 (1994)}

\lref\Baxn{Baxter, R.J.:
Eight-vertex model
in lattice statistics and
one-dimensional anisotropic
Heisenberg chain\semi
1. Some fundamental 
eigenvectors. Ann. Phys. (N.Y.) {\bf 76}, 1-24 (1973)\semi
2. Equivalence to
a generalized ice-type model. Ann. Phys. (N.Y.)
{\bf 76}, 25-47 (1973)\semi
3. Eigenvectors of the transfer matrix and Hamiltonian.
Ann. Phys. (N.Y.) {\bf 76}, 48-71 (1973)}

\lref\Factor{ Mussardo, G.: Off-critical statistical models,
factorized scattering theories and bootstrap program.
Phys. Rep. {\bf 218}, 215-379 (1992)}

\lref\Kane{Kane, C.L. and Fisher, M.P.A.:
Transmission through barriers and resonant tunneling in
an interacting one-dimensional electron gas.
Phys. Rev.  {\bf B46}, 15233-15262 (1992)}

\lref\Moon{Moon, K., Yi, H., Kane, C.L., Girvin, S.M. and
Fisher, M.P.A.: Resonant tunneling between quantum Hall
edge states. Phys. Rev. Lett. {\bf 27}, 4381-4384 (1993) }

\lref\FSLSN{Fendley, P., Ludwig, A.W.W. and Saleur H.:
Exact Conductance through point contacts in
the $\nu=1/3$ fractional quantum Hall effects.
Phys. Rev. Lett. {\bf 74}, 3005-3008 (1995) }

\lref\SalF{Fendley, P., Lesage, F. and Saleur H.:
A unified framework for the Kondo problem and
for an impurity in a Luttinger liquid.
Preprint USC-95-20, \#cond-mat/9510055 (1995)}

\lref\BLZn{Bazhanov, V.V., Lukyanov, S.L. and Zamolodchikov, A.B.: 
Integrable Structure of Conformal Field Theory III. The Yang-Baxter
Relations. To appear.}

\lref\MorsFesh{Morse, M.P. and Feshbach, H.:
Methods of theoretical physics.
New York, Toronto, London: McGraw-Hill book company, Inc. 1953 }

\lref\NeznayuN{Mumford, D.: Tata Lectures on Theta I, II.
Progress in Mathematics. Vol. {\bf 28}, 43.
Birkh$\ddot{{\rm a}}$user, Boston, Basel, Stuttgart 1983, 1984 }

\lref\Bern{Abramowitz, M. and Stegun, I.:
Handbook of mathematical function, New York:
Dover publications, Inc. 1970 }

\lref\SFN{Fendley, P. and Saleur, H.:
Exact perturbative solution of the Kondo problem.
Phys. Rev. Lett. {\bf 75},  4492-4495 (1995) }

\lref\Schmid{Schmid, A.:
Diffusion and localization in a dissipative
quantum system.
Phys. Rev. Lett. {\bf 51}, 1506-1509 (1983)}

\lref\leggett{Caldeira, A.O.
and Legget, A.J.: Influence
of dissipation on quantum tunneling in
macroscopic systems. Phys. Rev. Lett. 
{\bf 46}, 211-214 (1981)\semi
Caldeira, A.O.
and Legget, A.J.:
Path integral approach to quantum
Brownian motion.
Physica  {\bf A121}, 587-616 (1983)}

\lref\Callan{Callan, C.G. and Thorlacius L.:
Open string theory as dissipative quantum 
mechanics. Nucl. Phys. {\bf B329}, 117-138 (1990)}

\lref\Fisher{Fisher, M.P.A. and Zwerger, W.:
Quantum Brownian motion in a
periodic potential.  Phys. Rev.
{\bf B32}, 6190-6206 (1985)}

\lref\BLZNN{Bazhanov, V.V., Lukyanov, S.L. and Zamolodchikov, A.B.:
In preparation.}

\lref\FSN{ Fendley, P., Ludwig, A.W.W. and Saleur, H.:
Exact non-equilibrium transport through
point contacts in quantum wires and
fractional quantum Hall devices.
Phys. Rev. {\bf  B52} 8934-8950 (1995)}

\lref\Warn{Fendley, P., Saleur, H. and Warner, N.P.:
Exact solution of a massless scalar
field with a relevant boundary interaction.
Nucl. Phys. {\bf B430}, 577-596 (1994)} 

\lref\Frenk{Feigin, B. and Frenkel, E.:
Integrals of motion and quantum groups.
Proceeding of C.I.M.E. Summer School on
``Integrable systems and Quantum groups'', $\#$hep-th/9310022  }

\lref\Eguchi{Eguchi, T. and Yang, S.K.:
Deformation of conformal field theories and soliton
equations. Phys. Lett. {\bf B224}, 373-378  (1989)}

\lref\Stratonovich{Stratonovich, R.L.: Topics
in the Theory of Random Noise. Vol. 2,  
Chapter 9. New-York: Gordon and Breach  1967}

\Title{\vbox{\baselineskip12pt\hbox{CLNS 96/1405  }
\hbox{LPTENS 96/18}
\hbox{hep-th/9604044}}}
{\vbox{\centerline
{Integrable Structure of Conformal Field Theory II.}
\vskip6pt\centerline{ Q-operator and DDV equation}}}
\centerline{Vladimir V. Bazhanov$^1$\footnote{$^*$}
{e-mail address: Vladimir.Bazhanov@maths.anu.edu.au}, 
Sergei L. Lukyanov$^{2,4}$ \footnote{$^{**}$}
{e-mail address: sergei@hepth.cornell.edu}}
\centerline{ 
and Alexander B. Zamolodchikov$^{3,4}$ \footnote{$^\dagger$}
{On leave of absence from Department of Physics and Astronomy,
Rutgers University, Piscataway,
NJ 08855-0849, USA} \footnote{$^\&$}{Guggenheim 
Fellow} }
\centerline{$^1 $ Department of 
Theoretical Physics and Center of Mathematics}
\centerline{and its Applications, IAS, Australian National University, }
\centerline{Canberra, ACT 0200, Australia}
\centerline{and}
\centerline{Saint Petersburg Branch of Steklov Mathematical Institute,}
\centerline{Fontanka 27, Saint Petersburg, 191011, Russia}
\centerline{$^2$Newman Laboratory, Cornell University}
\centerline{ Ithaca, NY 14853-5001, USA}
\centerline{$^3$Laboratoire de Physique
Th\'eorique de l'Ecole Normale Sup\'erieure,}
\centerline{24 rue Lhomond - 75231 Paris Cedex 05, France}
\centerline{and}
\centerline{$^4$L.D. Landau Institute for Theoretical Physics,}
\centerline{Chernogolovka, 142432, Russia}

\Date{April, 96}
\eject

\centerline{{\bf Abstract}}

This paper is a direct continuation  of\   \BLZ\ 
where we begun the study of the integrable structures
in Conformal Field Theory.  We show here how to
construct the operators ${\bf Q}_{\pm}(\lambda)$ which act in 
highest weight Virasoro module and commute for different
values of the parameter $\lambda$. These operators appear to be the 
CFT analogs of the $Q$ - matrix of Baxter\ 
\Baxn , in particular they satisfy famous 
Baxter's ${\bf T}-{\bf Q}$ equation.
We also show that under natural assumptions 
about analytic properties of the operators ${\bf Q}(\lambda)$ as the
functions of $\lambda$ the Baxter's relation allows one to
derive the nonlinear integral equations
of Destri-de Vega (DDV)\ \dVega\ 
for the eigenvalues
of the ${\bf Q}$-operators. We then use the DDV equation
to obtain the  asymptotic expansions of the ${\bf Q}$ -
operators at large $\lambda$; it is remarkable that unlike the expansions
of the ${\bf T}$ operators of \ \BLZ, the asymptotic
series for  ${\bf Q}(\lambda)$
contains the ``dual'' nonlocal Integrals of Motion 
along with the local ones.  We also discuss an intriguing  relation
between the vacuum
eigenvalues of the ${\bf Q}$ - operators and 
the stationary transport properties in boundary sine-Gordon model.
On this basis we propose a number of new exact results about 
finite voltage charge transport through the point contact in quantum 
Hall system.

\vfil
\eject

\newsec{Introduction}
Existence of an infinite set of mutually
commuting local Integrals of Motion
(IM) is the characteristic feature of an
integrable quantum field theory
(IQFT). Therefore simultaneous diagonalization
of these local IM is the 
fundamental problem of IQFT. In the case of infinite-size
system this problem
reduces to finding mass spectrum and factorizable
S-matrix associated with 
IQFT; much progress in this direction has been made
during the last two decades (see e.g. \Factor\ for a review).
On the other hand, for a finite-size system 
(say, with the spatial coordinate
compactified on a circle of circumference $R$) 
this problem becomes highly 
nontrivial and so far its solution is
known to a very limited extent.  Most 
important progress here has been made with
 the help of so called Thermodynamic 
Bethe Ansatz (TBA) approach\ \YY ,\ \Zar.
TBA allows one
to find the eigenvalues associated 
with the ground state of the system 
(in particular the ground-state energy) in
terms of solutions of nonlinear integral equation 
(TBA equation). 
However it is not clear how the combination
of thermodynamic and relativistic 
ideas which is used in traditional
derivation of the TBA equation
can be extended to include the excited states.

The above diagonalization problem
is very similar to that treated in
solvable lattice models. In the lattice
theory very powerful algebraic and 
analytic methods of diagonalization of the Baxter's 
families of commuting 
transfer-matrices are known\ \Baxn,\  \Baxter;
these methods are further developed in Quantum
Inverse Scattering Method (QISM) \ \FadS,  \Kor .
Of course many IQFT can be obtained 
by taking continuous limits of solvable
lattice models and the method based 
on commuting transfer-matrices can be
used to solve these QFT. This is 
essentially the way how IQFT
are treated in the QISM. However, 
for many IQFT (notably, for most of IQFT defined as perturbed 
CFT \ABZ ) the associated solvable lattice models
are not known. Besides,
it seems to be conceptually important
to develop the above methods 
directly in continuous QFT, in particular,
to find continuous QFT versions 
of the Baxter's commuting transfer-matrices. 

This problem was addressed in our recent paper \ \BLZ\ 
where we concentrated 
attention on the case of
Conformal Field Theory (CFT), more specifically on
$c<1$ CFT. We should stress here that although the structure 
of the space of states and
the energy spectrum in CFT are relatively well 
understood the diagonalization
of the {\it full} set of the local  IM 
remains very nontrivial open problem. In
\ \BLZ\  we have constructed an infinite 
set of operator 
valued functions
${\bf T}_j (\lambda)$, where $j = 0, \frac{1}{2}, 1,
\frac{ 3}{2}, ... $ and 
$\lambda$ is a complex variable.
These operators (we will exhibit their 
explicit form in Sect.2) act invariantly in irreducible
highest weight Virasoro 
module ${\cal V}_{\Delta}$
and they commute between themselves for any 
values of $\lambda$, i.e.
\eqn\per{\eqalign{&{\bf T}_j (\lambda) :
\quad {\cal V}_{\Delta}  \to  
{\cal V}_{\Delta}\ ,\cr
&[{\bf T}_j (\lambda),\  {\bf T}_{j'} (\l')]=0\ .}}
The operators ${\bf T}_j (\lambda)$ are defined in terms of certain 
monodromy matrices associated with $2j+1$ dimensional representations 
of quantum  algebra $U_q (sl_2 )$\ \lref\Rks,  where
\eqn\kiuy{q = e^{i\pi\beta^2}\ .}
and $\beta$ is related to the Virasoro central charge as 
\eqn\central{c = 13-6\, \big(\beta^2+\beta^{-2}\big)\ .}
Evidently, the operators ${\bf T}_j (\lambda)$ are CFT versions of the 
commuting transfer-matrices of the Baxter's lattice theory. We 
will still call these operators ``transfer-matrices''
although the original
meaning of this
term\ \Baxter\  apparently is lost. As we have shown in \ \BLZ , 
in CFT the 
operators ${\bf T}_j (\lambda)$ enjoy particularly simple analytic 
properties, namely they are entire 
functions of $\lambda^2$ with an essential singularity at $\lambda^2 =
\infty$ and their asymptotic behavior
near this point is described in terms 
of the  local IM. Therefore the operators ${\bf T}_j (\lambda)$ can be 
thought of as the generating functions for the local IM since all the 
information about their eigenvalues is contained in the eigenvalues of 
${\bf T}_j (\lambda)$. The operators ${\bf T}_j (\lambda)$ are shown 
to obey the ``fusion relations'' which for any rational value of 
$\beta^2$ in \ \kiuy\  provide a finite system of functional equations
for the eigenvalues of these operators.  For the ground-state 
eigenvalue (in CFT it corresponds to a primary state) these functional 
equations turn out to be 
equivalent to the TBA equations; in general case they provide modified
TBA equations suitable for the excited states. Interesting but somewhat 
inconvenient feature of this approach is that the resulting TBA 
equations depend on $c$ in a very irregular manner (they depend on the 
arithmetic properties of the rational
number $\beta^2$) whereas the resulting
eigenvalues of ${\bf T}_{j} (\lambda)$ are
expected to be smooth functions 
of $c$. 

Another powerful method known in the lattice theory is based on the 
so-called ${\bf Q}$ -operator. This method was introduced by Baxter in 
his original study of 8-vertex model\ \Baxn.
One of its advantages is that it is 
not limited to the cases when $q$ is a root
of unity. In this paper (which 
is a sequel to \BLZ)
we introduce the analog of ${\bf Q}$-operator directly 
in CFT and study its properties. The ${\bf Q}$-operators (in fact we will 
define two ${\bf Q}$-operators, ${\bf Q}_{\pm}(\lambda)$) are defined
again as the traces of certain monodromy matrices, this time associated
with infinite-dimensional representations of so called ``$q$-oscillator 
algebra''. The operators ${\bf Q}_{\pm}(\lambda)$ obey the Baxter's
functional relation
\eqn\baxg{{\bf T}(\lambda) {\bf Q}(\lambda) = {\bf Q}(q\lambda) + 
{\bf Q}(q^{-1}\lambda)\ , }
where ${\bf T}(\lambda) \equiv {\bf T}_{1\over 2}(\lambda)$. This 
construction is presented in Sect.2 where also the most 
important properties of the ${\bf Q}$-operators are discussed.

In the lattice theory the Baxter's relation  \ \baxg\ is known to be a 
powerful tool for finding
the eigenvalues of the transfer-matrices\ \Baxter, the
knowledge about analytic properties of the ${\bf Q}$-operator being 
a key ingredient
in this approach.
Our construction of the ${\bf Q}$-operators as 
the traces of the 
monodromy matrices makes it natural to assume that they enjoy very simple 
analytic properties similar
to those of the operators  ${\bf T}_j (\lambda)$:  
up to overall power-like factors they are  entire 
functions of $\lambda^2$  with the following  asymptotic at 
$\lambda^2 \to -\infty$
\eqn\jubt{\log\  {\bf Q}_{\pm}(\lambda) \sim M\
(-\lambda^2)^{\frac{1}{2-2 \beta^2}}\ ,} 
where $M$ is a constant 
(which will be actually calculated in Sect.3).
Using these properties we show that
the eigenvalues of the ${\bf Q}$-operators 
satisfy the  Destry-de Vega (DDV) equation\ \dVega .
This is done in Sect.3. The DDV equation can be solved exactly in the 
limit $\Delta \to +\infty$, where $\Delta$ is the Virasoro highest weight
in \per , and we analyze in the
Sect.3 the vacuum eigenvalues of the operators 
${\bf Q}(\l)$ in this limit.
In Sect.4 we further study the properties of the 
${\bf Q}$ - operators and formulate our basic conjectures about their 
analytic characteristics.
The exact asymptotic expansions of the ${\bf Q}$ and 
${\bf T}$ operators at $\l^2 \to \infty$ are proposed here.
We observe that unlike the
asymptotic expansion of ${\bf T}(\l)$ the large $\l^2$  expansion 
of ${\bf Q}(\l)$ contains both local
and nonlocal IM and that the operators
${\bf Q}(\l)$ obey remarkable duality
relation with respect to the substitution
$\beta^2 \to  \beta^{-2}$.
Although the results of this section have somewhat 
conjectural status we
support them by explicit study of the eigenvalues 
of the ${\bf Q}$ operators at the 
``free fermion point'' $\beta^2 = 1/2$. In Sect.5
we discuss the relation of the 
${\bf Q}$ operators to the characteristics of
stationary non-equilibrium states
in so called boundary sine-Gordon model\ \Warn\  (see also \ \GZ); these
states attracted lately much attention
in relation to the finite-voltage current
through the 
point contact in a quantum Hall system \ \Kane,\  \Moon ,\ \FSN,\ \SalF .
Possible directions of further studies
are discussed in Sect.6.

\newsec{ The {\bf Q}-operators}

In this section we will introduce the 
operators ${\bf Q}_{\pm}(\lambda)$ which 
satisfy \ \baxg . We start with a brief 
review of the definitions and 
results of \ \BLZ .

Let $\varphi(u)$ be a free chiral 
Bose field, i.e. the operator-valued function
\eqn\vars{\varphi(u) = iQ + i P u + \sum_{n\neq 0}
{{a_{-n}}\over {n}}\  
e^{inu}\ . }
Here $ P,  Q$ and $ a_n, n = \pm 1, \pm 2, 
.$ are operators which satisfy
the commutation relations of the Heisenberg
algebra
\eqn\osc{[ Q, P] = {i\over 2}\ \beta^2 ; \qquad [ a_n ,  a_m] = 
{n\over 2}\ \beta^2\  \delta_{n+m, 0}\ .}
with real $\beta$.
The variable  $u$ is interpreted as a complex coordinate on $2D$ 
cylinder of a circumference $2\pi$. As follows 
from \ \vars\  the field $\varphi (u)$ is a 
quasi-periodic function of $u$, i.e.
\eqn\perka{\varphi(u+2\pi) = \varphi(u) + 2\pi i P\ .}
Let ${\cal F}_p$ be the Fock space, i.e. the space
generated by a free action 
of the operators $a_n$ with $n < 0$ on the vacuum
vector $\mid p \rangle$ which satisfies
\eqn\kuyhg{\eqalign{
&a_n \mid p \rangle = 0\ , \quad {\rm  for} \quad n > 0\ ;\cr
& P\mid p \rangle = p\mid p \rangle\ .}}

The composite field 
\eqn\nuyt{- \beta^2 T(u) =
:\varphi' (u)^2: + (1 - \beta^2)\varphi''(u) + 
{{\beta^2}\over 24}  }
is called the energy-momentum 
tensor; it is periodic function of $u$
and its Fourier modes 
\eqn\tens{L_n = \int_{-\pi}^{\pi}{du\over{2\pi}}\ 
\bigg[ \ T(u) + {c\over 24}\ \bigg] \ e^{inu} }
generate the Virasoro algebra with
the central charge \  \central \ \Fei,\ \DotsFat. It is well 
known that for generic values 
of the parameters $\beta$ and $p$ the Fock
space ${\cal F}_p$ realizes an irreducible
highest weight Virasoro module
${\cal V}_{\Delta}$ with the 
highest weight $\Delta$ related to $p$ as
\eqn\juyt{\Delta =\Big(\frac{p}{\beta}\Big)^2 +\frac{c-1}{24}\ .}
For particular values of these parameters, when null-vectors appear
in ${\cal F}_p$, ${\cal V}_{\Delta}$ is obtained from ${\cal F}_p$ by 
factoring out all the invariant subspaces. The space
\eqn\mju{{\hat {\cal F}}_{p}=
\oplus_{n=-\infty}^{\infty} {\cal F}_{p+n{\beta^2}} }
admits the action of the exponential fields 
\eqn\eip{V_{\pm}(u) = :e^{\pm 2 \varphi (u)}:\ .}
Also, let $E, F$ and $H$ be a canonical
generating elements of the algebra
$U_q \big(sl(2)\big)$\ \RKSA, i.e.
\eqn\kjhg{[H,E]=2E,
\qquad [H,F]=-2F, \qquad [E,F]={{q^{H}-q^{-H}}\over {q -
q^{-1}}}\ , }
where $q$ is given by \kiuy . Let $j$ be a non-negative integer or 
half-integer number. 
We denote $\pi_j$ an irreducible $2j+1$ dimensional matrix
representation of
$U_q \big(sl(2)\big)$
so that $E_j \equiv \pi_j (E),\ 
F_j \equiv \pi_j (F)$ and  $H_j \equiv \pi_j (H)$
are $(2j+1)\times (2j+1)$
matrices which satisfy the relations \ \kjhg .

The ``transfer-matrices'' ${\bf T}_j (\lambda),\ 
j=0, {1\over 2}, 1, {3\over 2},
 ... $ are defined as  \ \BLZ
\eqn\nhy{\eqalign{
&{\bf T}_j (\lambda) = 
tr_{\pi_j} \bigg[ e^{2\pi i P H_j } {\cal P}\exp\big(
\lambda \int_{0}^{2\pi} K_j (u) du\big)\bigg] \cr\equiv
tr_{\pi_j} \bigg[ e^{2\pi i P H_j }& \sum_{n=0}^{\infty} \lambda^n 
\int_{2\pi \geq u_1 \geq u_2 \geq ... \geq u_n \geq 0}
K_j (u_1)K_j (u_2) ...
K_j(u_n)\  du_1 du_2 ... du_n \bigg]\ .}}
Here 
$$K_j (u) = V_{-}(u)\  q^{{H_j}\over 2} E_j + V_{+}(u)\ 
q^{-{{H_j}\over 2}} F_j $$
and ${\cal P}$ denotes the operator 
ordering along the integration path. 
Obviously, 
\eqn\juytaa{{\bf T}_0(\lambda) = {\bf I} \ ,}
where ${\bf I}$ is the identity operator. Although the
exponentials in \ \nhy\  
act from one component of the sum \ \mju\ 
to another it is easy to see that the
operators \ \nhy\  invariantly act 
in the Fock space ${\cal F}_p$. Another 
obvious property of ${\bf T}_j$ is that
these operators are in fact the 
functions of $\lambda^2$ as
the traces of the odd-order terms in \ \nhy\  vanish.

As is explained in \ \BLZ\  the 
operators \nhy\  form the commuting family, 
i.e. they satisfy the relations \ \per .
They also obey the ``fusion relations''
\eqn\fus{{\bf T}(\lambda)
{\bf T}_j (q^{j+{1\over 2}}\lambda) = 
{\bf T}_{j-{1\over 2}}
(q^{j+1}\lambda) + {\bf T}_{j+{1\over 2}}(q^{j}\lambda)\ .}
Together with \ \juytaa\  this relations allow one 
to express recurrently any of
the ``higher-spin'' operators
${\bf T}_j (\lambda)$ with $j = 1, {3\over 2}, 
2, ...$ in terms of the basic one
\eqn\kiujhg{{\bf T}(\lambda) \equiv {\bf T}_{1\over 2}(\lambda)\ .}
In the case $j={1\over 2}$ it is easy to evaluate the traces in
\ \nhy\  and obtain the power-series expansion
\eqn\knhdy{{\bf T}(\lambda) = 2 \cos (2 \pi P) + \sum_{n=1}^{\infty}
\lambda^{2n}\  { {\bf G}}_{n}\ ,}
where the coefficients define an infinite set
of basic nonlocal IM
\eqn\mjui{\eqalign{
{{\bf G}}_{n} = q^n& \int_{2\pi\geq u_1 \geq u_2 \geq ... \geq u_{2n} 
\geq 0} \big(\ 
e^{2 i\pi P}\  V_{-}(u_1) V_{+}(u_2) V_{-}(u_3) ... 
V_{+}(u_{2n} ) +\cr &
e^{-2 i\pi P}\  V_{+}(u_1) V_{-}(u_2) V_{+}(u_3)
. V_{-}(u_{2n})
\ \big)
\ du_1 du_2 ... du_{2n}\ .}}
It follows from
\ \per\  that these nonlocal IM commute among themselves
$$[{\bf  G}_{n},
{{\bf G}_{m}}] = 0\ .$$

The operator-product expansion 
\eqn\hdgf{ V_{+}(u) V_{-}(u') = (u-u')^{-2\beta^2}\big(\, 
1+O(u-u')\, \big) \ ,\ \  u-u'\to 0 } 
shows that the expressions \ \nhy\ 
and\  \mjui\  can be taken literally only if
\eqn\oiuy{0<\beta^2 < {1\over 2}\ , }
for otherwise the integrals in \ \mjui\ 
diverge. In what
follows we will call the region\ \oiuy\ 
the Semi-classical
Domain (SD). In fact one can define the 
operators ${\bf G}_{n}$ outside the
SD by analytic continuation in $\beta^2$.
A convenient way to do that is to
transform the ordered integrals in\ \mjui\ 
to contour integrals. For example
${{\bf G}_1}$ can be written as 
\eqn\lkjht{\eqalign{
&{ {\bf G}_1} =
\big(q^2-q^{-2}\big)^{-1}\  \int_{0}^{2\pi}d u_1 \int_{0}^{2\pi}
d u_2\
\bigg\{\  \big(q e^{-2\pi i P} - q^{-1} e^{2\pi i P}\big)\times
\cr &
V_{-}(u_1 + i0) V_{+}(u_2 - i0) +
\big(q e^{2\pi i P} - q^{-1} e^{-2\pi i P}\big)\ 
V_{+}(u_1 + i0) V_{-}(u_2 - i0)\ \bigg\} \ . }}
Obviously, \ \lkjht\ 
does not contain divergent integrals and for 
$\beta^2$ in  SD
coincides with the
ordered integral in \ \mjui . Similar representation 
exists for higher ${\bf  G}_{n}$,
and thus the operator ${\bf T}(\lambda)$ 
can be defined outside  SD through 
\ \knhdy . Of course the operators 
${\bf G}_{n}$ and ${\bf T}(\lambda)$
obtained this way exhibit singularities 
at $\beta^2 = \beta_{n}^2$
\eqn\kiuy{\beta_{n}^2 = {{2n-1}\over 2n},
 \qquad n = 1, 2, 3, ... \  ,}
where the integrals \ 
\mjui\  develop logarithmic divergences. In order to define
the operators ${\bf T}_j (\lambda)$ at the singular 
points \ \kiuy\  some 
renormalization is needed but we do not discuss it here
(see however our analysis of the case $\beta^2 =1/2$ in Sect.4).

As is mentioned in the Introduction, the power series
\ \knhdy\  defines 
${\bf T}(\lambda)$ as an entire
function of $\lambda^2$ with an 
essential singularity at $\lambda^2 \to \infty$.
Its asymptotic  expansion
near this essential singularity 
can be expressed in terms of local IM
as
\eqn\loexp{\log {\bf T}(\lambda)
\simeq m\  \lambda^{\frac{1}{1-\beta^2}}
\ {\bf I} -\sum_{n=1}^
{\infty}\ C_n\  
\lambda^{\frac{1-2n}{1-\beta^2}}
\ {\bf I}_{2n-1}\ , }
where ${\bf I}_{2n-1}$ is the basic 
set of commutative local IM as defined in\  \BLZ ; the 
operators ${\bf I}_{2n-1}$ can be written as the integrals
\eqn\loint{{\bf I}_{2n-1} = \int_{0}^{2\pi}\frac{du}{2\pi }\ 
T_{2n}(u) \ , }
where the local densities $T_{2n}(u)$
are particular normal ordered polynomials 
of\ $ \partial_u\varphi  (u)$,..., 
$\partial^{2n}_u\varphi  (u)$
(see  \BLZ\  for details and for our convention
about the normalization of ${\bf I}_{2n-1}$).
The numerical coefficients $m$ and
$C_n$ in the expansion\  \loexp\ 
will be calculated exactly in Sect.3. 
As follows from\  \per\  all these local 
IM commute with the nonlocal IM ${\bf G}_n$.

As is known\ \jap,\ \Eguchi,\  \Frenk,
the local IM ${\bf I}_{2n-1}$ defined in \ \BLZ\  do not 
change under the substitution $\beta^2 \to \beta^{-2}$, 
if we simultaneously make the  replacement
\ $  \varphi(u)\to\ \beta^{-2}\varphi(u)$
\eqn\kiuy{{\bf I}_{2n-1}\big\{\beta^2,\varphi(u)\big\}=
{{\bf I}}_{2n-1}\big\{\beta^{-2},\beta^{-2}\varphi(u)\big\}\ . }
Evidently the nonlocal IM
do change and so there exists an infinite set of 
``dual'' nonlocal IM ${\tilde {\bf G}}_{n}$
which are obtained from \  \mjui\  by just this 
substitution, i.e.
\eqn\juytr{\eqalign{{\tilde {\bf G}}_{n} = 
{\tilde q}^n& \int_{2\pi\geq u_1 \geq u_2 \geq ... \geq u_{2n} 
\geq 0} \big(
\ {\tilde q}^{2P}\  
U_{-}(u_1) U_{+}(u_2) U_{-}(u_3) ... U_{+}(u_{2n}) +\cr
&\ {\tilde q}^{-2P}\ U_{+}(u_1) U_{-}(u_2) U_{+}(u_3) ... 
U_{-}(u_{2n})\ \big)\ 
du_1 du_2 ... du_{2n }\  ,}}
where
$$U_{\pm}(u) = :e^{\pm \frac{2 }{\beta^2} \varphi(u)}: $$
and
\eqn\kiuyoi{{\tilde q} = e^{i\frac{\pi}{\beta^2}}\ .}
Of course if $\beta^2 < 2$ the analytic 
continuation described above is needed
to define ${\tilde {\bf G}}_{n}$. However
it is possible to check that
the operators ${\tilde {\bf G}}_{n}$ thus defined 
commute among themselves and commute with all the
nonlocal IM ${\bf G}_n$ and local IM ${\bf I}_{2m-1}$
\eqn\kjuh{
[{\tilde {\bf G}}_{n},
{\tilde {\bf G}}_{m}]=[{\tilde {\bf G}}_{n}, {\bf G}_{m}]= 
[{\tilde {\bf G}}_{n}, {\bf I}_{2m-1}] = 0\ .}

Now we are in position to define 
the ${\bf Q}$-operators and to describe 
their basic properties. 
Let ${\cal E}_+ , {\cal E}_- $ and ${\cal H}$ be 
operators which satisfy
the commutation relations of so called ``q-oscillator algebra'',
\eqn\osc{ q\,
{\cal E}_{+}{\cal E}_{-} - q^{-1} {\cal E}_{-}{\cal E}_{+} = 
{1\over {q-q^{-1}}}\ , \qquad [{\cal H}, {\cal E}_{\pm}] = 
{\pm}2{\cal E}_{\pm}}
and let $\rho$ be any representation 
of this algebra such that the trace
\eqn\shdg{Z(p) = tr_{\rho} [\ e^{2\pi i p {\cal H}}\ ]\ }
exists for complex $p$
belonging to the upper half plane, $\Im  m\  p > 0$. 
Then one can define two operators
\eqn\kjhg{{\bf A}_{\pm} (\lambda) =  Z^{-1}(\pm P) tr_{\rho}
\big[\  e^{\pm 2 i\pi P {\cal H}} {\cal P} 
\exp \big( \lambda \int_{0}^{2\pi} du( V_{-}(u)
q^{\pm\frac{{\cal H}}{ 2}} 
{\cal E}_{\pm} +
V_{+}(u)q^{\mp\frac{{\cal H}}{2}} {\cal E}_{\mp})\big) \ \big]\ ,}
where again the symbol ${\cal P}$ denotes the operator ordering along 
the integration domain. As we are interested in the action of these 
operators in ${\cal F}_p$ the operator $P$ in \ \kjhg\ 
can be substituted
for its eigenvalue $p$. Strictly speaking the definition
\ \kjhg\  makes
sense only if $\Im  m\  p > 0$
for ${\bf A}_{+}$ and if $\Im  m\  p < 0$ for 
${\bf A}_{-}$. However these operators can be defined for all
complex $p$ (except for some set of singular
points on the real axis) by 
analytic continuation in $p$.
Then it is easy to see that ${\bf A}_{-}$ can be 
obtained from ${\bf A}_{+}$ by substitution
\eqn\lje{ P \to -P, \qquad \varphi(u) \to -\varphi(u)\ .}

The operators \ \kjhg\  can be written as the power series 
\eqn\kiuyt{
{\bf A}_{\pm}(\lambda) = 1+
\sum_{n=1}^{\infty} \sum_{\{\sigma_i=\pm 1\}
\atop\sigma_1+\cdots+\sigma_{2n}=0} 
\lambda^{2n}\  a_{2n}(\sigma_1,\ldots,\sigma_{2n}|\pm P) 
\ J_{2n}(\mp \sigma_1, \mp \sigma_2, \ldots, \mp \sigma_{2n})\  ,}
where
$$J_{2n}(\sigma_1, \ldots, \sigma_{2n}) =q^n\
\int_{2\pi\ge u_1\cdots u_{2n}\ge0}
V_{\sigma_1}(u_1) V_{ \sigma_2}(u_2)\cdots V_{ \sigma_{2n}}(u_{2n})
\ du_1\ldots du_{2n} $$
and 
\eqn\hytr{a_n(\sigma_1,\ldots,\sigma_{2n}|P)=Z^{-1}(P)\ tr_{\rho}
\bigg(e^{2\pi i P {\cal H}}
{\cal E}_{\sigma_1}
{\cal E}_{\sigma_2}\cdots{\cal E}_{\sigma_{2n}}\bigg)\ .}
It is easy to see that the coefficients \ \hytr\ 
are completely determined
by the commutation relations\
 \osc\   and the cyclic property of the trace
and so the operators\ \kjhg\ do not
depend on the particular choice of 
the representation ${\rho}$.
The operator coefficients in the power series
\ \kiuyt\  can be expressed in terms of the basic nonlocal IM \ \mjui .
It is not 
difficult to calculate first few terms of this series
\eqn\oikjn{\eqalign{
{\bf A}_{\pm}&(\lambda) = 1-
\lambda^2\  {{ {\bf G}_1}\over 4 \sin a \sin(a\pm x)} 
-\l^4\ \biggl\{\ 
\frac{{\bf G}_2}{4 \sin 2a \sin(2a\pm x)}-\cr
&\frac{{\bf G}^2_1}{16
\sin a \sin 2a \sin(a\pm x) \sin(2a\pm x)}\ \biggr\}-
\l^6\ \biggl\{\ 
\frac{{\bf G}_3}{4  \sin 3a \sin(3a\pm x)}+\cr &
\ \ \ \ \ \ \ \   
\frac{{\bf G}^3_1-4\ {\bf G}_1{\bf G}_2
\big(\sin a\sin(a\pm x)+\sin 2a \sin(2a\pm x)\big)}
{64 \sin a \sin 2a  \sin 3a \sin(a\pm x) \sin(2a\pm x)
\sin(3a\pm x)}\  \biggr\}+O(\l^8)\ ,}} 
where
$$x = 2\pi P , \qquad a= \pi \beta^2 \ .$$
For further references it is convenient to introduce an alternative  set 
of nonlocal IM defined 
as coefficients in the expansion
\eqn\Aser{
\log {\bf A}_+(\l)=-\sum_{n=1}^\infty
\  y^{2n}\ {\bf H}_{n}\   ,}
where
\eqn\ylam{ 
y=\beta^{-2} \Gamma(1-\beta^2)\  \l\ .}
These coefficients are, of course, algebraically dependent on those
in \ \mjui . For example, 
\eqn\htwo{
{\bf H}_1=
{\beta^4\ \Gamma(\beta^2)\over
4\pi\ 
\Gamma(1-\beta^2)\ \sin(2\pi P+\pi\beta^2)}\  {{\bf G}_1}\ .}
Define also a new set of ``dual'' nonlocal IM
$\widetilde{{\bf H}}_{n}$ 
\eqn\hdual{
\widetilde{{\bf H}}_{n}\big\{\beta^2,\varphi(u)\big\}=
{{\bf H}}_{n}\big\{\beta^{-2},\beta^{-2}\varphi(u)\big\}\ ,}
obtained from\  ${\bf H}_{n}$ by the  analytic continuation, as described
above.

The operators ${\bf Q}_{\pm}(\lambda)$ are  defined as
\eqn\kiuyt{{\bf Q}_{\pm}(\lambda) = \lambda^{\pm 2P/{\beta^2}} 
{\bf A}_{\pm} (\lambda) \ .}
Like the operators ${\bf T}_j (\lambda)$ above the operators
\ \kiuyt\ 
act in a Fock space ${\cal F}_p$
\eqn\kjhnm{{\bf Q}_{\pm}(\lambda) : 
\quad {\cal F}_p\to
{\cal F}_p\ .}
The operators $ {\bf Q}_{\pm}(\lambda)$ exhibit remarkable
properties. Here we simply list some of them leaving the proofs to
the other paper.

{\it i.} The operators ${\bf Q}_{\pm}(\lambda)$ commute among themselves 
and with all the operators ${\bf T}_j (\lambda)$,
\eqn\kon{[{\bf Q}_{\pm}(\lambda), {\bf Q}_{\pm}(\lambda')] = 
[{\bf Q}_{\pm}(\lambda),
{\bf T}_j (\lambda')] = 0\ .}

{\it ii.} The operators ${\bf Q}_{\pm}(\lambda)$ satisfy
the equation\  \baxg ,
i.e.
\eqn\lkjh{{\bf T}(\lambda){\bf Q}_{\pm}(\lambda) = {\bf Q}_{\pm}(q\lambda)
+ {\bf Q}_{\pm}(q^{-1}\lambda)\ .}
The equation \ \baxg\ 
can be thought of as the finite-difference analog of
the second order differential 
equation so we expect it to have two linearly independent
solutions. As ${\bf T}(\lambda)$ is a single-valued function of 
$\lambda^2$, i.e. it is periodic 
function of $\log \lambda^2$, the operators
${\bf Q}_{\pm}(\lambda)$ are just
two ``Bloch-wave'' solutions to the
equation \ \baxg . The operators
${\bf Q}_{\pm}(\lambda)$ satisfy the 
``quantum Wronskian'' condition
\eqn\wron{{\bf Q}_{+}(q^{1\over 2}\lambda)
{\bf Q}_{-}(q^{-{1\over 2}}\lambda) - 
{\bf Q}_{+}(q^{-{1\over 2}}\lambda){\bf Q}_{-}(q^{1\over 2}\lambda) = 
2i\  \sin ( 2\pi P )\ .}

{\it iii.} The ``transfer-matrices''
${\bf T}_j(\lambda)$ can be expressed in
terms of ${\bf Q}_{\pm}(\lambda)$ as
\eqn\mnhy{2i\ \sin(  2\pi P  )
\ {\bf T}_j (\lambda) = {\bf Q}_{+}(q^{j+{1\over 2}}\lambda)
{\bf Q}_{-}(q^{-j-{1\over 2}}\lambda) -
{\bf Q}_{+}(q^{-j-{1\over 2}}
\lambda){\bf Q}_{-}(q^{j+{1\over 2}}\lambda)\ .}
In view of this equation the operators ${\bf Q}_{\pm}(\lambda)$ appear
more fundamental then the transfer-matrices 
${\bf T}_j(\lambda)$. We will see more support to this idea below.

Let us just briefly sketch the derivation \ \BLZn\ 
of the functional relation
of this section.
The main idea is to consider more general $\bf T$-operators
${\bf T}^+_j(\l)$ defined as in \ \nhy, 
but associated with the infinite
dimensional representation $\pi^+$ of $U_q\big(sl(2)\big)$
with arbitrary (complex)
highest weight $2j$.
Note that  if $j$ takes  a non-negative integer or 
half-integer value then the
matrices $\pi_j^+(E)$, $\pi_j^+(F)$ and $\pi_j^+(H)$\ 
have a
block-triangular form with two diagonal blocks, one  equivalent 
to the $(2j+1)$-dimensional representation $\pi_j$ and the other
being the highest weight representation $\pi_{-j-1}^+$. In this
way we obtain the following simple relation
\eqn\simple{
{\bf T}_j^+(\l)={\bf T}_j(\l)+{\bf T}_{-j-1}^+(\l)\ ,
\qquad j=0,1/2,1,3/2,\ldots\ .
}
Next, the operator ${\bf T}^+_j(\l)$ enjoy the following remarkable
factorization property
\eqn\factora{
2 i \sin(2\pi P)\ {\bf T}^+_j(\l)={\bf Q}_+(q^{j+\hf}\, \l)\,
{\bf Q}_-(q^{-j-\hf}\, \l)\ ,}
which is proved explicitly by using decomposition properties of the
tensor product of
two representations of the $q$-oscillator algebra (the latter are also
representations of the Borel sub-algebra of 
$U_q\big(\widehat{sl(2)}\big)$ with
respect to the co-multiplication from 
$U_q\big(\widehat{sl(2)}\big)$. Then
the functional relations \ \wron, 
\mnhy\  trivially follow from \simple\  and \ \factora\ 
above,
while the remaining relations  \ \kon,\ \lkjh\ 
are simple
corollaries of these two.

Although our
considerations here were specific to the continuous theory
similar results hold for $Q$-matrix of  the lattice
theory as well.  This
is quite obvious since
the structure of the functional equations is determined
merely by the decomposition properties of products of representations of
$U_q\big(\widehat{sl(2)}\big)$
associated with the monodromy matrices. The
details of these calculations will be given in \ \BLZn.

The operators ${\bf Q}_{\pm}(\lambda)$ take particularly simple form
when applied to the space ${\cal F}_p$ with $2p$ equals some integer
which 
we denote $N$.
In these cases
the ``quantum Wronskian'' \ \wron\ 
is equal to zero  and the solutions\ ${\bf A}_{+}(\l)$ \ and \ 
${\bf A}_{-}(\l)$\ 
\kiuyt\
coincide . For $2p = N$ all the coefficients \ \hytr\  are readily 
calculated
\eqn\lkjhg{a_{2n}(\sigma_1, \ldots , \sigma_{2n}|\ N/2)
= (q - q^{-1})^{-2n} \ ,}
and the operators \ \kiuyt\ 
can be written as
\eqn\kdiyb{{\bf A}_{\pm}(\lambda)\Big|_{p = N/2} = 
\sum_{n=0}^{\infty}
\  {\mu^{2n}\over ({n!})^2}\   
\ q^n\ {\cal P}\biggl\{\  \Big[ \int_{0}^{2\pi}
\ \frac{du}{2\pi}V_{+}(u)\ 
\Big] ^n \Big[ \int_{0}^{2\pi}\ \frac{du}{2\pi}
V_{-}(u)
\ \Big] ^n
\ \biggr\}\Big|_{p = N/2}
\  , }
where
\eqn\kijhy{\mu =-i \frac{\pi \lambda}{\sin (\pi\beta^2)}\ }
and, as before,
the symbol ${\cal P}$ denotes  ``$u$-ordering'', i.e. the operators 
$V_{\pm}(u)$ with greater $u$ are placed to the left. For $p=N/2$ 
the exponentials $V_{\pm}(u)$ are $2\pi$-periodic functions of $u$ 
and so the integration contours in \kdiyb\ close. The series \ \kdiyb\ 
is closely related to the boundary state in so called Boundary 
Sine-Gordon model (with the bulk mass equal zero)\Warn.
In particular,
its vacuum-vacuum matrix element \ $A_{\pm}^{(vac)}(\lambda )\ $, 
defined as
\eqn\lokjh{
{\bf A}_{\pm}(\lambda)
\mid p \rangle=A_{\pm}^{(vac)}(\lambda)\mid p \rangle\ ,}
coincides for $p=N/2$ with the one-dimensional Coulomb gas partition 
function
\eqn\mjuyt{\eqalign{
&A^{(vac)}(\lambda)|_{p= N/2} \equiv Z_{N}(\mu)=
\sum_{n=0}^{\infty}\  \frac{\mu^{2n}}{(n!)^2}
\   \int_{0}^{2\pi}\frac{du_1}{2\pi} 
\ldots \int_{0}^{2\pi}
\frac{du_n}{2\pi} \int_{0}^{2\pi}\frac{dv_1}{2\pi} 
\ldots \int_{0}^{2\pi}\frac{dv_n}{2\pi}\times\cr
& e^{iN\sum_{i=1}^{n} (v_i-u_i)}\ 
{\prod_{i\not=j}\Big|4\sin\big(\frac{u_i-u_j}{2}\big)
\sin\big(\frac{v_i-v_j}{2}\big)\Big|^{2\beta^2}}
{\prod_{i,j}
\Big|2\sin\big(\frac{u_i-v_j}{2}\big)\Big|^{-2\beta^2}}\ .}}
As was shown in \ \FSLS\  the series 
\ \mjuyt\   defines an entire function of $\mu^2$
with the asymptotic behavior $\log  Z_{N}(\mu) \sim 
const\ (\mu^2 )^{1\over {2-2\beta^2}}$. In fact, it is easy
to show that this result implies
not only to the vacuum eigenvalue  \ \lokjh\ 
but also to any matrix element
of the operator ${\bf A}$. We conclude that for  $p=N/2$ the 
operators ${\bf A}_{\pm}(\lambda)$ are 
entire functions of $\lambda^2$
and they enjoy the asymptotic form
\eqn\ytr{\log \ {\bf A}_{\pm}(\lambda) \sim  M\  
(-\lambda^2)^{1\over {2-2\beta^2}}\ ,\ \ \ \ \ \l^2\to-\infty\   .}
We should stress again that the
relations \ \kdiyb\  and \ \mjuyt\  hold only
for  $2p=N$. For non-integer $2p$ they
do not hold. 
Nonetheless we find it natural to assume that the above analytic 
properties of the operators ${\bf A}_{\pm}(\lambda)$ as the functions
of $\l^2$ hold
for any $p$.
In the following Sections we use the functional equation
\ \baxg\
together with
this analyticity assumption to
derive various asymptotic expansions for the
operators ${\bf A}_{\pm}(\l)$.

\newsec{Destri-de Vega equation} 

Now let us turn to the eigenvalue problem for the operators
${\bf Q}_{\pm}(\lambda)$
\eqn\kiuytr{{\bf Q}_{\pm}(\lambda)\mid  \alpha \rangle
 = Q_{\pm}^{\alpha}(\lambda) \mid  \alpha \rangle\ ,}
where $\mid  \alpha \rangle \in {\cal F}_p$. In this section we 
consider only the case of $\beta^2$ in SD\ \oiuy . Let us concentrate 
attention on one of the operators \kiuyt , say 
${\bf Q}(\lambda) \equiv {\bf Q}_{+}(\lambda)$; the problem
for ${\bf Q}_{-}(\lambda)$ can be solved then by the substitution 
\lje . 

As follows from\  \lkjh\ 
any eigenvalue $Q(\lambda)$ of ${\bf Q}(\lambda)$
satisfies the Baxter's functional equation
\eqn\TQQ{ 
{T}(\l) {Q}(\l)= {Q}(q \l) + {Q}(q^{-1} \l)\ ,}
where $T(\l)$ is the corresponding eigenvalue of the operator
${\bf T}(\l)$. Denoting $A(\l)$ the eigenvalue of the operator
${\bf A}_{+}(\l)$ in\  \kiuyt\  one can rewrite \TQQ\ as
\eqn\TAp{
 { T}(\l) { A}(\l)=e^{2\pi i p }{ A}(q \l) + 
e^{-2\pi i p}{ A}(q^{-1} \l)\ .}
Although this equation looks as just a relation between two unknown
functions in fact it imposes severe restrictions on these functions
provided their analytic properties are known\foot{In lattice 
theory the equation \TAp\ (which appears there
``decorated'' with some 
non-universal factors in the 
right hand side) leads to the Bethe Ansatz
equations which completely 
determine the eigenvalues \ \Baxter.}. Motivated by the
relation to the Coulomb gas partition function\  \mjuyt\ 
discussed in Sect.2
we accept here the following 
assumptions about analytic properties of
the eigenvalues $A(\l)$.

Let $\beta^2$ be restricted to  SD and let $\Im m\  p = 0$.
Then

(i) {\it Analyticity.} The functions $A(\l)$ and $T(\l)$ are entire
functions  of the complex variable \  $\,\l^2\,\quad$\ 
\foot{In fact
one can prove this property for $T(\l)$, see  \BLZ.}.

(ii) {\it Location of zeroes.} Zeroes of the function\ $A(\l)$ in the
$\l^2$-plane  are either real or occur in complex 
conjugated pairs. For
any given eigenvalue $A(\l)$ 
there are only finite number of complex 
or real negative zeroes.
Real zeroes accumulate toward $+\infty$ in the 
$\lambda^2$. For the vacuum eigenvalues 
all zeroes are real and if $2p >
-\beta^2$ they are all positive.

(iii) {\it Asymptotic behavior.} 
The leading asymptotic behavior of $A(\l)$ 
for large $\l^2$
is given by \ytr\ with some constant $M$.

If $0<\beta^2<\half$\  an entire 
function $A(\l)$ with the asymptotic 
behavior\  \ytr\  is completely 
determined by its zeroes $\l_{k}^2$,
$k=0,1,\ldots$\ ; it
can be represented by a convergent product
\eqn\sd{A(\l)=\prod_{k=0}^{\infty}
\bigg(1-{\l^2\over\l_k^2}\bigg)\ ,}
where  
the normalization condition
\eqn\normak{ A(0) = 1\ }
is taken into account.

Actually, one important remark about the normalization
\kiuyt,\ \normak\ of
the operators ${\bf A}$ must be made here. For given
eigenvalue $A(\l)$ the positions of its zeroes
$\l_k$ depend on $p$ and for special values of
this parameter one of these zeroes,
say $k_0$, can happen to be at zero, 
$\l_{k_0} = 0$. For example, for the vacuum eigenvalue
$A_{+}^{(vac)}(\l)$ this happens at $2p = -\beta^2 -
n; \quad n = 0, 1, 2, ...$, as brief inspection of 
\oikjn\ and (3.21) below shows. Obviously, at these
values of $2p$ the normalization of the operators 
${\bf A}$ as in \ \kiuyt,\  \normak\ is not suitable as under this
normalization all the terms in the power series 
$A(\l) = 1 + \sum_{n=1}^{\infty} \l^{2n}\, A_n$ except the 
first one would diverge. Of course this formal 
singularity can be eliminated by renormalization 
of $A$ with appropriate $p$-dependent factor. 
As our results below are not sensitive to this
subtlety, in what follows we use the normalization
\ \kiuyt,\  \normak\ (which we find technically very convenient)
and just assume that $p$ does not take these ``dangerous''
values (i.e. that that all $\l_k \neq 0$).

Introduce the  function
\eqn\al{
a(\l)=e^{4\pi i p}{A(\l q)\over A(\l q^{-1})}\ .}
Setting $\l^2=\l_k^2$ in  \TAp\ and remembering that $T(\l)$ has no 
singularities at finite $\l^2$ one obtains the Bethe-Ansatz type 
equations for the positions of the zeroes $\l_k^2$
\eqn\BAe{a(\l_k)=-1\ .}
This is an infinite set of transcendental equations
for the infinite set
of unknowns $\l_k^2$. 
It can be transformed into a coupled system of a 
single non-linear integral 
equation and a finite set of equation \BAe\ only 
for those zeroes  which do not belong to the positive real axis in the 
$\l^2$-plane. This system 
is known as Destry-de Vega (DDV) equation\ \dVega.
In the Appendix A we show 
that under the assumptions (i)-(iii) given above
the infinite set of equations \BAe\ is equivalent to
\eqn\ddv{\left\{\matrix{\displaystyle
i \log a(\theta)=-\frac{2\pi  p}{\beta^2}
+2M  {\rm cos}\frac{\pi\xi}{2}\ e^\theta
+i{\sum_a}'{\rm log}\      S(\theta-\theta_a)
-2 G\ast \Im m\, \log\big(1+a(\theta-i0)\big)\cr\cr
\phantom{i \log}a(\theta_a)=-1\ ,\hfill\cr}\right.}
where we have used new variables 
\eqn\variab{
\l=e^{\frac{\theta}{1+\xi}}\ ,\qquad 
\l_a=e^{\frac{\theta_a}{1+\xi}}\ ,
\ \qquad \beta^2=\frac{\xi}{1+\xi}\  }
and $\l_a^2$ denote zeroes of 
$A(\l)$ lying outside the positive real axis of 
$\l^2$.
The star in \ddv\ denotes the convolution
\eqn\ewq{ A\ast B(\theta)=
\int_{-\infty}^{\infty}d\theta' A(\theta-\theta')  B(\theta')}
and
\eqn\jxmki{S(\theta)=
\exp\Biggl\{-i
\int_{0}^{\infty}\frac{d\nu}{ \nu}\
\sin( \nu \theta)\ {\sinh\big(\, \pi\nu(1+ \xi) /2\, \big)
\over  \cosh(\, \pi \nu/2\, )
\sinh(\,  \pi\nu\xi/2\, )}\ \Biggr\}\ ,}
$$G(\theta)=\delta(\theta)+\frac{1}{2\pi i}\ \partial_{\theta}
{\rm log}\  S(\theta)\ .$$
Note that the function\ $S(\theta)$\ coincide with the soliton-soliton
scattering amplitude for the Sine-Gordon model \Zkvadrat.
The equation\  \ddv\ 
for the vacuum eigenvalues 
(in fact, more general equation obtained by replacing $e^\theta$ in the
r.h.s. by $2\sinh \theta$) was originally derived from the Bethe ansatz
equations associated with XXZ lattice model by taking appropriate 
continuous limit which leads to the sine-Gordon field theory \ \dVega.
Here we
obtain (the ``mass-less'' version of) this equation directly in continuous 
field theory, bypassing any direct reference to the lattice theory.

Given a  solution of 
\ddv\  the function\ ${\rm  log}A(\l)$\ can be
calculated as\ \dVega
\eqn\Asol{{\rm log}A(\l)=-i\int_{C_\nu}d\nu \  
\frac{g(\nu)}{\cosh\pi\nu/2\sinh \pi\nu\xi/2}\
\big(- \lambda^2\, 
\big)^{i\nu(1+\xi)/2}\ ,}
where the function $g(\nu)$
is defined as 
\eqn\defg{g(\nu)=\int_{-\infty}^{+\infty}\frac{d\theta}{2\pi}\ 
\Im m\, \log\big(1+a(\theta-i0)\big)\  e^{-i\nu\theta}\ .}
The integration
contour $C_\nu$ in \Asol\   goes  along the line\ $\Im m \ 
\nu=-1-\epsilon$ with arbitrary small positive $\epsilon$.

It is instructive to study the vacuum eigenvalue $A^{(vac)}(\l)$ 
of  ${\bf A}(\l)$, defined by \lokjh , 
in the limit $p \to +\infty$. Note that the parameter $p$ enters
the DDV equation
\ddv\  exactly the same way the external gauge field (which couples 
to the soliton charge)
appears in the sine-Gordon DDV equation \ \Zamt.
The DDV
equation \ddv\  is greatly simplified in the limit
$p \to +\infty$.
According to the assumption (ii) above  
all zeroes $\l_k^2$ in
this case are real and positive and so one can drop the term  
$ i{\sum_a}'{\rm log}\   S(\theta-\theta_a)$\
in the r.h.s. of the integral equation \ddv.
Moreover, in the limit
$p \to +\infty$ this  equation  reduces to a linear equation of 
Winer-Hopf type\    (see Appendix A for some details)
\eqn\lopaa{-\frac{\pi  p}{\beta^2}
+M  {\rm cos}\frac{\pi\xi}{2}\ e^\theta
-\int_{-\infty}^{B(p)}\frac{d\theta'}{2\pi i}\ \partial_{\theta}
{\rm log}\ S(\theta-\theta')\ \Im m
\log\big(1+a^{(vac)}(\theta')\big)=0\ , }
provided one assumes that 
\eqn\juio{B(p)={{1+\xi}\over 2}\ \log\ \l_0^2 \sim
const\  \log p\ ,\qquad p\to+\infty\ ,}
where $\l_0^2$ is a minimal   of the zeroes $\l_k^2$.

The equation \lopaa\ can be solved
by the standard technique\ \MorsFesh. As a result one
obtains
\eqn\kot{g^{(vac)}(\nu)|_{p\to+\infty}\sim
\frac{i p\ \sqrt\pi }{4   \nu}\ 
\frac{\Gamma\big(1-i\nu(1+\xi)/2\big)}
{\Gamma(3/2-i\nu/2)\ \Gamma(1- i\nu\xi/2)}\ e^{i\delta\nu}\
\big(\,  \l_0^2\, \big)^{-i\nu(1+\xi)/2}\ ,}
where $g^{(vac)}(\nu)$ is related to $a^{(vac)}(\t)$ through
\defg   
\eqn\hdyf{
\big(\,  \l_0^2\, \big)^{(1+\xi)/2}\ \Big|_{p\to+\infty}\sim  
\frac{\Gamma(\frac{\xi}{2})\   \Gamma(\frac{1}{2}
-\frac{\xi}{2})}
{ \sqrt\pi\ M}
\ e^{\delta} \  p\  ,}
and
$$2\delta=  (1+\xi)\ {\rm \log}(1+\xi)
-\xi\ {\rm log}\xi\ .$$
Note that \ \hdyf\  supports the assumption\ \juio . 
The equation\ \Asol\  gives
\eqn\kidg{\eqalign{
&\ \ \ \ \ \ \ \ \ \ 
\ \ \ \ \ \ \ \ \ \ \ \ \ \ \ {\rm log} A^{(vac)}(\l)|_{p\to+\infty}\sim
\cr
&-\frac{p}{2\pi^{\frac{3}{2}}\xi}\ 
\int_{C_{\nu}}\frac{d\nu}{\nu^2}\ 
\Gamma\big(1-i\nu (1+\xi)/2\big)\ \Gamma(1+i\nu\xi/2)\ \Gamma(
-\frac{1}{2}+i\nu/2)\ e^{i\delta\nu}\
\Big(- \frac{\l^2}{\l^2_0}\, \Big)^
{i\nu (1+\xi)/2}\ ,}}
where integration contour\ $C_{\nu}$\  go along the line
\ $\Im  m\ \nu=-1-\epsilon$. 
The integrand\ $\sigma(\nu)$\ 
in \kidg\  has the simple  asymptotic behavior in whole
complex $\nu$-plane except positive imaginary axis
$$\sigma(\nu)\sim
\frac{p\ \sqrt{\beta^2}}
{ \sinh\big(\pi\nu(1+\xi)/2\big)}\
 \Big(- \frac{\l^2}{\l^2_0}\, \Big)^
{i\nu (1+\xi)/2}\  , \qquad \nu \to
\infty$$
and we can work out the  integral\ \kidg\ for\ $|\l^2|<|\l^2_0|$
\   by closing
the integration contour\ $C_{\nu}$\  at
the infinity in the lower  half plane\ $\Im m\   \nu<-1$\ 
and calculating the residues of the poles at 
\ $\nu=-2 i n/(1+\xi), \ 
n=1,2,...\ .$ The result can be written in the form 
\eqn\aas{
\log A^{(vac)}(\l)|_{p\to+\infty}=
-\sum_{n=1}^{\infty} \ y^{2n}\ H^{(vac)}_{n}|_{p\to\infty}\ ,}
where $y$ is given by\  \ylam\  and the above residue calculations 
give the coefficients $H^{(vac)}_{n}|_{p\to\infty}$ explicitly. 
According to\  \Aser\ 
this coefficients coincide with the large $p$ asymptotic  
of the vacuum eigenvalues
of the nonlocal IM  ${\bf H}_{n}$. In particular, we find
\eqn\htwoas{
H_{1}^{(vac)}|_{p\to+\infty}\sim
\frac{\beta^4\,\Gamma(\beta^2)\Gamma(1/2-\beta^2)}{2\sqrt\pi}
\ p^{2\beta^2-1}\ .}
On the other hand an exact expression 
for $H_1^{(vac)}$ for all values of $p$
is known explicitly from \htwo
\eqn\htwoex{
H_1^{(vac)}=
\frac{\beta^4\,\Gamma(\beta^2)\Gamma(1-2\beta^2)}{\Gamma(1-\beta^2)}\
\frac{\Gamma(\beta^2+2 p)}{\Gamma(1-\beta^2+2 p)}\ ,}
where we have used the known vacuum eigenvalue 
\eqn\qvac{G_{1}^{(vac)}=
\frac{4\pi^2\ \Gamma(1-2\beta^2)}{\Gamma(1-
\beta^2-2p)\ \Gamma(1-
\beta^2+2p)}\ .}
of the nonlocal IM ${\bf G}_1$\ \BLZ\  and \ \htwo .
Comparing the asymptotic of
\htwoex\ with \htwoas\  one obtains exactly the 
coefficients  \ $M$\  
\ which enters the leading asymptotic behaviors \  \ytr\   
of 
the operators ${\bf A}(\l)$:
\eqn\Mexact{M=\frac{\Gamma(\frac{\xi}{2})  \Gamma(\frac{1}{2}
-\frac{\xi}{2})}{\sqrt\pi}\ 
\Big(\Gamma\big(1-\beta^2\big)\Big)^{1+\xi}\ . }
Calculating further coefficients in \aas\  we find the large $p$
asymptotic of the eigenvalues $H_n^{(vac)}$
\eqn\hnas{H_{n}^{(vac)}|_{p\to+\infty}\sim 
\frac{\Gamma(n\beta^2 )\
\Gamma\big(-\frac{1}{2}+n(1-\beta^2) \big)}{2\ \sqrt\pi n!}\ 
\big(\,  \beta^2\, \big)^{2n}\ \ p^{1-2n+2n\beta^2}\ .}
It is not obvious at all how this asymptotic can be obtained 
directly from the definition of nonlocal IM $H_n$ in terms of 
the ordered integrals  \mjui\  through \knhdy \  and \Aser. 

The large $p$ asymptotic of vacuum eigenvalues of 
the dual nonlocal IM \hdual\   is obtained from 
\hnas\ by the substitution $\beta^2\to \beta^{-2}$,
$p\to \beta^{-2}\ 
p$
\eqn\hdas{\widetilde{H}_{n}^{(vac)}|_{p\to+\infty}\sim
\frac{\Gamma(n\beta^{-2} )\
\Gamma\big(-\frac{1}{2}+n(1-\beta^{-2}) \big)}{2\ \sqrt\pi n!}\
\big(\,  \beta^2\, \big)^{-\frac{2n}{\beta^2}-1}\ 
\ p^{1-2n+\frac{2n}{\beta^2}}\ .}
For completeness we also present here the large $p$ asymptotic of the 
eigenvalues of the local IM 
\eqn\limas{
I_{2n-1}^{(vac)}|_{p\to+\infty}\sim
\big(\, \beta^{2}\, \big)^{-n}\  p^{2n}\ ,}
which follow directly from the definition of  
the operators ${\bf I}_{2n-1}$, see \ \BLZ. 
Note that this asymptotic hold for any
eigenvalues (not just the vacuum 
ones) of the local IM.

Let us now consider the large $\l$ behavior of \kidg. To obtain
large $\l$ asymptotic expansion of $A^{(vac)}(\l)$ one can close 
the integral in \kidg\ in the upper half-plane $\Im m\  \nu\ge-1$. Then
\kidg\  can be represented as a sum of infinitely many terms associated
with the residues of the poles of the integrand located
at $\Im m\  \nu\ge-1$,
plus the integral over
the large circle. Note that the contribution of the
large circle in fact diverges. Correspondingly, the sum of
the residues gives only asymptotic series expansion for $A^{(vac)}(\l)$,
with zero radius of convergence. This asymptotic expansion has the form
\eqn\Aaspi{\eqalign{A^{(vac)}(\l)\simeq
C^{(vac)}\big(\beta^2,p\big)&\ \big(- y^2\, \big)^{-\frac{p}{\beta^2}}\ 
\exp\Biggl\{\sum_{n=0}^{\infty} B_{n}\
\big(-y^2\,
\big)^{\frac{1-2 n}{2-2\beta^2}}\  I^{(vac)}_{2n-1}\ 
\Biggr\}\times\cr 
&\exp\Biggl\{-\sum_{n=1}^{+\infty}\  (-1)^n\
\big(- y^2\, \big)^
{-\frac{ n}{\beta^2}}\ 
\widetilde{H}^{(vac)}_{n}
\ \Biggr\}\ .}}
where the variable $y$ is related to $\l$ as in \ylam ,
\eqn\Cpi{C^{(vac)}\big(\beta^2,p\big)\big|_{p\to+\infty}\sim
\big(\, \beta^2\, \big)^{-\frac{2p}{\beta^2}}
\, \Big(\ \frac{2p}{ e}\, \Big)^{\frac{2p}{\beta^2}-2p},  }
and 
\eqn\Bdef{
B_n=\frac{(-1)^{n+1}}{2\sqrt\pi\ (1-\beta^2)\ n!}\ 
\Gamma\Big( \frac{2n-1}{2-2\beta^2} \Big)\
\Gamma\Big(\frac{2n-1}{2-2\beta^{-2}}\Big)\ 
\big(\, \beta^2\, 
\big)^{\frac{n\beta^2+n-\beta^2}{\beta^2-1}}\ .}
Strictly speaking, here
we have derived the expansion \Aaspi\ 
in the limit $p \to +\infty$, so the quantities
$I_{2n-1}^{(vac)}$ and ${\tilde H}_{n}^{(vac)}$
in \Aaspi\ denote the $p \to +\infty$ 
asymptotic of the corresponding vacuum
eigenvalues given by \  \limas\ and\   \hdas .
However in the next section we adopt additional 
analyticity assumptions about 
the operators ${\bf A}(\l)$ which allow for
derivation of the same
asymptotic expansion \Aaspi\ for arbitrary $p$.
Moreover, it is natural 
to expect that the
expansion \Aaspi\ holds not
only for the vacuum eigenvalues 
but for the whole operator ${\bf A}(\l)$; in that case 
of course the vacuum 
eigenvalues $ I_{2n-1}^{(vac)}$ and ${\tilde  H}_n ^{(vac)}$
in \Aaspi\ must be replaced by the operators 
${\bf I}_{2n-1}$ and ${\tilde {\bf H}}_n$
themselves.

\newsec{Conjectures: exact asymptotic expansions and duality}

Consider the following 
operator-valued function of the complex variable $\nu$ 
\foot{In writing (4.1) we make a technical assumption
that all zeroes $\l_k$ of all eigenvalues of ${\bf A}$
are away from zero, see the remark in Sect.3.} 
\eqn\defpsi{\eqalign{
{\bf \Psi}(\nu) =&
\frac{2\sqrt\pi\
\Big(\Gamma(1-\beta^2)\Big)^{-i\nu(1+\xi)}}{\Gamma\big(i\nu\xi/2\big)
\ \Gamma\big(-1/2+i\nu/2\big)\ 
\Gamma\big(-i\nu(1+\xi)/2\big)}\times\cr 
&\int_{-\infty}^0 {d\l^2 \over \l^2}
\big(-\l^2\, \big)^{-i\nu(\xi+1)/2}\ \log{\bf A}(\l)\ .}}
The function ${\bf \Psi}(\nu)$ provides a continuous
set of IM parameterized by the variable $\nu$. Indeed, the operators 
${\bf A}(\l)$ with different values
of $\l$ as well as their arbitrary 
linear combinations commute among themselves and with all local and
nonlocal IM. 
As we shall see below
the function ${\bf \Psi}(\nu)$ is remarkable in many 
respects, in particular, it can be thought as an analytic continuation
of the
local IM ${\bf I}_{2n-1}$ to arbitrary complex values of their index $n$. 

The integral \defpsi\ converges
only for $2\beta^2-2<\Im m\,\nu<-1$,
however the
definition of ${\bf \Psi}(\nu)$ 
can be extended to the whole $\nu$-plane 
by means of the analytic
continuation. The latter 
 can be done in various ways. For instance, 
using the product representation \ \sd\  which holds 
for any eigenvalue $A(\l)$ of the operator ${\bf A}(\l)$ one
can write the corresponding eigenvalue of ${\bf \Psi}(\nu)$ as 
a generalized Dirichlet series of the form\foot{The definition
\defpsi\ contains an ambiguity in the choice of the branches
of the logarithm. In (4.2)\ this translates into the choice of the
phases of the roots $\l_k^2$.  This ambiguity, however,
does not affect the
following arguments. For definiteness one may assume that 
$|\arg \l_k^2|<\pi$ when $\l_k^2$ lies outside the negative real axis
of $\l^2$, while for real negative $\l_k^2$ the corresponding entry in
the sum (4.2)  is replaced by $\cosh\big(\pi \nu (1+\xi)/2\big)\ 
|\l_k|^{-i\nu(1+\xi)}$.}
\eqn\Dir{ \Psi(\nu)=2\sqrt\pi\
\frac{\Gamma\big(i\nu(1+\xi)/2\big)
\ \Big(\Gamma(1-\beta^2)\Big)^{-i\nu(1+\xi)}}{\Gamma\big(i\nu\xi/2\big)
\ \Gamma\big(-1/2+i\nu/2\big)}
\ {\sum_{k=0}^{\infty}} 
\ \big(\, \l_k^2\, \big)^{-i\nu(1+\xi)/2}\ .}
For $0<\beta^2<1/2$ this 
 series   converges  absolutely  for $\Im m\ \nu<-1$ so that \Dir\
defines an analytic function of $\nu$ in the half plane 
 $\Im m\ \nu<-1$ which can be then analytically continued to the whole
complex $\nu$-plane by the  standard technique
of the analytic continuation of Dirichlet
series \ \NeznayuN.
We expect that the function ${\bf \Psi}(\nu)$ defined in
this way enjoy the following remarkable analytic properties

{\bf Conjecture 1.}
{\it
For real\ $p$ \ and \ $0<\beta^2\leq\half$ the function 
\  ${\bf \Psi}(\nu)$\ 
is an entire function of the complex variable\ $\nu$. }

These simple analytic properties makes
${\bf \Psi}(\nu)$ extremely convenient for studying eigenvalues 
of the ${\bf A}$ and ${\bf T}$ operators. 
Converting the integral transform in \defpsi\ one expresses
${\bf A(\l)}$ as
\eqn\apsi{\eqalign{
{\rm log}\  {\bf A}(\l)=
-\frac{i}{4\pi^{\frac{3}{2}}}
\int_{C_\nu}\frac{d\nu}{\nu}\
&\Gamma\big(1-i\nu (1+\xi)/2\big)\ \Gamma(i\nu\xi/2)\ \Gamma(
-1/2+i\nu/2)\times\cr
&\Big(\Gamma(1-\beta^2)\Big)^{i\nu(1+\xi)}\ 
{\bf \Psi}(\nu)\
\ \big(-\l^2\, \big)^{i\nu (1+\xi)/2}\  ,}}
where   the integration  contour\ $C_{\nu}$ is the same as in\ \Asol .
At the same time, the corresponding  eigenvalue of the transfer matrix
\ ${\bf T}(\l)$\  reads
\eqn\tlam{
{\bf T}(\l)=
{\bf \Lambda}(q^{\frac{1}{2}}\l)
+{\bf \Lambda}^{-1}(q^{-\frac{1}{2}}\l)\ ,}
where 
\eqn\lampsi{\eqalign{
{\rm log}\ {\bf \Lambda}(\l)=2\pi i P+
\int_{C_\nu}\frac{d\nu}{\nu}\
&\frac{\Gamma\big(1-i\nu (1+\xi)/2\big)\ \Gamma(-1/2+i\nu/2)}
{2\sqrt\pi\  \Gamma(
1-i\nu\xi/2)}\times\cr
&\Big(\Gamma(1-\beta^2)\Big)^{i\nu(1+\xi)}\
{\bf \Psi}(\nu)\
\ \big(-\l^2\, \big)^{i\nu (1+\xi)/2}\ .}}
The values of the function $ \Psi(\nu)$ at special points 
on the imaginary $\nu$-axis (where the Gamma-functions in \apsi\ and
\lampsi\ display poles)
are of particular interest. For example, 
it is not difficult to see that the values
$\Psi\big(-2in(1-\beta^2)\,\big)$, $n=1,2,\ldots,\infty$,
are related to the eigenvalues of the nonlocal IM ${\bf H}_{n}$.
Using \sd\ and \ \Aser\ 
one obtains for  the latter
\eqn\heig{
H_{n}=n^{-1}\ 
\Big(\beta^{-2}\Gamma(1-\beta^2)\Big)^{-2n}
\ \sum_{k=0}^{\infty}\ \l_k^{-2n}\ .}
Then it follows from \Dir\ that
\eqn\psih{
{\bf \Psi}\big( -2in(1-\beta^2)\, \big)=
\frac{2\ \sqrt\pi n!\ \big( \,  \beta^2\, \big)^{-2n}}
{\Gamma(n\beta^2 )\
\Gamma\big(-\frac{1}{2}+n(1-\beta^2) \big)}\ {\bf  H}_{n}\ .}
For other special values on the  imaginary $\nu$-axis 
we will adopt the following

{\bf Conjecture 2.} {\it For real\ 
$p$ and $0<\beta^2\leq\half$
the operator\  ${\bf \Psi}(\nu)$\  has the following special values 
on the imaginary $\nu$-axis
\eqn\psii{
{\bf \Psi}\big(\, (2 n -1) i\, \big)=
\big(\,  \beta^2\, \big)^{n}\ {\bf I}_{2n-1}\ ,
\ \ n=0,1,\ldots \ ,} 
\eqn\psiir{
{\bf \Psi}\big(\, 2in(\beta^{-2}-1)\, \big)=
\frac{2\ \sqrt\pi n!\ \big(\,  \beta^2\, \big)^{\frac{2n}{\beta^2}+1} }
{\Gamma(n\beta^{-2} )\
\Gamma\big(-\frac{1}{2}+n(1-\beta^{-2}) \big)}\ 
{{\bf \tilde H}}_{n}\ ,
\ \ n=1,2,\ldots\  ,}
\eqn\psiiq{{\bf \Psi}(0)=P\ ,}
where ${\bf I}_{2n-1}$
and $\widetilde{{\bf H}}_{n}$ denote the
local IM \ \loint\  and the dual nonlocal 
IM \ \hdual\  respectively.}

Note that \psii\ with $n=0$ reads
\eqn\triv{
{\bf \Psi}(-i)={\bf I}_{-1}\equiv {\bf I}\ ,}
where  {\bf I} is the identity operator.
The relations\  \psii-\psiiq\  together with
the conditions\ \kiuy ,  \hdual\   makes it natural to assume that 
the operator\ ${\bf \Psi}$\ satisfy the following  duality condition:

{\bf Conjecture 3.}
\eqn\kiuyaa{{\bf \Psi}\{\nu, \beta^2,\varphi(u)\}=
\big(\, \beta^2\, )^{1-i\nu}\ {{\bf \Psi}}\{\nu, \beta^{-2},
\beta^{-2}\varphi(u)\}\ . }

An initial motivation for the above  conjectures came from the
study of the large $p$ asymptotic of the vacuum eigenvalues
of ${\bf A}(\l)$
in  the
previous section. Indeed, comparing \kidg\ and \apsi\ one obtains
\eqn\psias{
\Psi^{(vac)}(\nu)|_{p\to+\infty}\sim p^{1-i\nu}\ ,}
which with an account of \ \limas\  and 
\ \hdas\  obviously satisfy all the statements of
the Conjectures 1-3 \foot{Like in \limas\ , the
asymptotic\ \psias \ holds for all
eigenvalues of ${\bf \Psi}$, not just the vacuum ones.}. 
Further motivations and justifications of the
these conjectures are discussed below.

We are now ready to derive exact asymptotic expansions of the
eigenvalues for large $\l^2$. This is achieved by calculating the
integrals \apsi\ and \lampsi\
as formal sums over residues in the upper half plane
$\Im m \, \nu\ge-1$. 
Using  \psii\  one thus has from \lampsi\ for $\l^2\to\infty$ 
\eqn\Lamas{{\rm log}\ {\bf \Lambda}(\l)\simeq
i\ m\ {\bf I}\ \big(-\l^2\, \big)^{\frac{1}{2-2\beta^2}}
-i\sum_{n=1}^{\infty}(-1)^n\  C_{n}\
\ \big(-\l^2\, \big)^{\frac{1-2n}{2-2\beta^2}}\  
 \l^{(1-2 n)(1+\xi)}\  
{\bf I}_{2n-1}\ ,}
where
\eqn\mexact{m=
\frac{2\sqrt\pi\  \Gamma(\frac{1}{2}
-\frac{\xi}{2})}
{   \Gamma(1-\frac{\xi}{2})}\
\Big(\Gamma\big(1-\beta^2\big)\Big)^
{1+\xi} }
and
\eqn\ytr{C_n=\frac{\sqrt\pi (1+\xi)}{  n!}\  \big(\, \beta^{2}\,
\big)^n\  
\frac{\Gamma\big((n-\frac{1}{2})(1+\xi)\big)}
{\Gamma\big(1+(n-\frac{1}{2})\xi\big)}\
\Big(\Gamma(1-
\beta^2)\Big)^{-(2n-1)(1+\xi)}\  .}
It follows then that
the large  $\l$  asymptotic
behavior of the operator\ ${\bf T}$\ is given
by \ \loexp , where
the numerical coefficients $m$ and $C_n$ are given
by the formulas \ \mexact\ and \ \ytr .
The asymptotic expansion \ \loexp\ holds 
for
\eqn\lamdom{
-\pi<\arg\,\l^2<\pi\ .}
Similarly 
one can get the asymptotic expansion for the
operator\ ${\bf A}(\l)$\ \apsi . Define the operator 
\eqn\Cdef{{\bf C}=
\big(\, \beta^2\, \big)^{-\frac{2P}{\beta^2}}
\ \Big( 2
\ e^{i\partial_{\nu}{\rm log }
{\bf\Psi}(0)-1}\Big)^{\frac{2P}{\beta^2}-2P}\ ,}
which does not depend on the  spectral parameter\  $\l$.
It is also convenient to use the variable $y$ given by \ \ylam\ 
instead of the
variable $\l$ and exhibit all arguments of\ ${\bf A}(i y)$:
$${\bf A}\big\{iy,\beta^2,\varphi(u) \big\}\equiv
\exp\Big(-\sum_{n=1}^{\infty}
\ (-1)^n\  \big(\, y^2\, \big)^{n}\ 
{\bf H}_{n}\ \Big)\ .$$
Then, from \apsi\ one has
for $|y|\to \infty$, \ $-\pi<arg\  y^2<\pi$
\eqn\Aas{\eqalign{{\bf A}\big\{iy,\beta^2,\varphi(u)\big\}\simeq
& {\bf C}\big\{\beta^2,\varphi(u)\big\}\ 
\big( \, y^2\, \big)^{-\frac{P}{\beta^2}}\
\exp\Biggl\{\, \sum_{n=0}^{\infty} B_{n}\
(\, y^2\, )^{\frac{1-2 n}{2-2\beta^2}}
\ {\bf I}_{2n-1}
\, \Biggr\}\times\cr
& {\bf A}\big\{i y^{-\frac{1}{\beta^2}}, \beta^{-2},
\beta^{-2}\varphi(u)\big\}\ ,}}
where the coefficients $B_n$ are given by \Bdef\ 
and
\eqn\Adual{
{\bf A}\big\{iy^{-\frac{1}{\beta^2}},
\beta^{-2},\beta^{-2}\varphi(u)\big\}
\equiv\exp\Big(-\sum_{n=1}^{\infty}
\ (-1)^n\
\big(\,  y^2\,
\big)^{-\frac{n}{\beta^2}}\ 
\widetilde{{\bf H}}_{n}\ \Big)\ .} 
The explicit form of the operator  ${\bf C}$\ in \Aas\ 
is not determined by the above calculations. For the vacuum
eigenvalue its $p\to+\infty$ asymptotic is given by \Cpi. 
Moreover, one can show that\ \FSLS 
\eqn\Czer{C^{(vac)}(\beta^2,0)=\sqrt{\beta^2}\ .}
We have the following conjecture about the exact form of this coefficient
for all values of $p$ and
$0<\beta^2\leq\frac{1}{2}$ 
\eqn\kiof{C^{(vac)}(\beta^2,p )=\sqrt{\beta^2}\ 
\frac{\Gamma\big(1+2p\beta^{-2}\big)}
{\Gamma\big(1+2p\big)}\ .}

The formula for the asymptotic expansion \Aas\ is essentially equivalent 
to the  Conjectures 1,2 given above. In fact, the coefficients in front
of powers of $\l$ in \Aas\ are in one-to-one correspondence with the
values \psii\  while any singularity of ${\bf \Psi}(\nu)$ in the upper
half plane would bring in some additional terms in \Aas.

The asymptotic expansions \loexp\ and \Aas\ are in remarkable agreement 
with the numerical calculation through the (modified) TBA equation. 
We postpone a detailed description of these calculations to a separate 
publication \ \lref\BSZNN\ 
but just mention some the results here.
In \ \BLZ\  we calculated numerically a few coefficients in \loexp\ for 
two vacuum states ($\Delta=-1/5$ and $\Delta=0)$
in the ${\cal M}_{2,5}$ CFT ($c=-22/5$)
and found an excellent agreement with the corresponding exact eigenvalues 
of the local IM given explicitly in \ \BLZ\  (up to
${\bf I}_7$ inclusive). 
Another numerical result concerning the part of \Aas\ containing  the
dual  nonlocal  IM is mentioned in Sect.5. 

As an additional support to our conjectures
consider the the case 
$$\beta^2=\half\ ,$$
where the eigenvalues of ${\bf T}$
and ${\bf A}$-operators
can be calculated explicitly. Moreover, the eigenvalues of the local IM 
can be independently found using the fermionic representation.
The value  $\beta^2=\half$ does not lie in SD\ \oiuy\ 
therefore the results of Sect.2 do not apply directly. In particular, 
the definitions 
\nhy\ and \kjhg\  for ${\bf T}(\l)$ and ${\bf A}(\l)$
requires a renormalization since 
 the nonlocal IM ${\bf G}_{2n}$ in \knhdy\  and \oikjn\ 
 diverge logarithmically at $\beta^2=\half$.
It turns out  that this renormalization affects the functional
equation \TAp .
To see this consider the expressions 
 for the vacuum eigenvalues to within the first order in $\l^2$ 
\eqn\Tbh{
T^{(vac)}(\l)=2\cos(2\pi
p) +\l^2\ G_1^{(vac)}+O(\l^4)\ ,}
\eqn\Abh{
A^{(vac)}(\l)=1-\frac{\l^2}{4\cos(2\pi p)}\ 
\Big(\ G^{(vac)}_1-2\pi^2 \sin(2\pi p)\ \Big)
+O(\l^4)\ ,}
where
\eqn\gdefaa{ G_1^{(vac)}=2 \cos(2\pi p)\ \Big(
\ 2{\cal  C}-
\pi\psi\big(\half-2p\big)-\pi\psi\big(\half-2p\big)\ \Big)\ ,}
$\psi(x)=\partial_x \log\Gamma(x)$ 
is the logarithmic derivative of the gamma-function
and ${\cal C}$
is a (non-universal) renormalization constant depending on the 
ultraviolet cutoff. A simplest way to obtain these expressions 
is to set $\beta^2=\half-\epsilon$,
$\epsilon\to0$ in \htwoex\ and \qvac;
that gives an analytic regularization 
of the divergent integrals with the value
of ${\cal C}$ 
\eqn\Cval{
{\cal C}={\pi\over2\epsilon}+\pi \psi(1)\ .}
With the above 
accuracy in $\l^2$ the eigenvalues 
\Tbh\ \Abh\ satisfy a  ``renormalized'' 
functional equation
\eqn\TAr{T(\l)A(\l)=e^{2\pi i p-i\pi^2\l^2}A(q\l)
+e^{-2\pi i p+i\pi^2 \l^2}A(q^{-1}\l)\ ,}
where $q=\exp(i\pi/2)$. 
Using the lattice 
regularization (i.e., considering discrete approximations to the
${\cal P}$-exponents in  \nhy\  and \ 
\kjhg\  and then tending the number of
partitions to infinity)
one can show that the functional equation \TAr\ is,
in fact, exact in the sense that it is valid for arbitrary  eigenvalues
$T(\l)$ and $A(\l)$ to all order in $\l^2$. 
 
The functional equation \TAr\  completely determine 
the eigenvalues $T(\l)$ and
$A(\l)$ provided one assumes them to be entire functions of
$\l^2$.
One can obtain
\eqn\Tsigma{
T(\l)=T^{(vac)}(\l)\,\prod_{k=1}^L F(i\l,p,n^+_k,n^-_k)\ 
F(i\l,-p,n^-_k,n^+_k)\ ,}
\eqn\Asigma{
A(\l)=A^{(vac)}(\l) \prod_{k=1}^L F(\l, p,n^+_k,n^-_k)\ ,}
where 
\eqn\Fdef{
F(\l,p,n^+,n^-)={(2p-n^-+\half-\pi\l^2)(2p+n^+-\half)\over
(2p+n^+-\half-\pi\l^2)(2p-n^-+\half)}}
and  $n^{\pm}_1 , n^{\pm}_2 ,\cdots, n^{\pm}_L$ are two
finite sequences 
of non-negative integers, $1\le n^{\pm}_1<\cdots<n^{\pm}_L$,
which uniquely specifies certain vector in the Fock space
${\cal F}_{p}$\foot{The basis of this kind in the
bosonic Fock space is  
usually referred to as ``fermionic basis''\ \MKLas,\ \McCoy.}; 
this vector has the Virasoro weight
\eqn\ksust{\Delta\{p;n^+_1,...,n^+_L;n^-_1,...,n^-_L\}=
\Delta^{(vac)}(p)+\sum_{k=1}^L(n^+_k+n^-_k-1)\ .}
For the vacuum eigenvalues $L=0$ and 
\eqn\Tvbh
{T^{(vac)}(\l)=\frac{2\pi\  e^{2{\cal C}\l^2}\ }
{\Gamma\big(1/2+2p+\pi\l^2\big)\ 
\Gamma\big(1/2-2p+\pi\l^2\big)}\ .}
\eqn\Avbh{A^{(vac)}(\l)=e^{-{\cal C}\l^2}
\frac{\Gamma\big(2p+1/2\big)}
{\Gamma\big(2p+1/2-\pi\l^2\big)}\ .}
It follows then from \Dir\ that
\eqn\psif{\eqalign{
\Psi(\nu)&=2^{i\nu-1}\ 
(i\nu-1)\ 
\Big\{\ \zeta\big(i\nu,2p+\half\big)+E_L(i\nu,p)\ \Big\}\ ,\cr
E_L(s,p)&=\sum_{k=1}^{L} \Big\{\ \big(2p+\half-n^-_k\big)^{-s}-
\big(2p-\half+n^+_k\big)^{-s}\ \Big\}\ ,}}
where $\zeta(s,\alpha)$ is the
generalized zeta function defined as 
the analytic continuation of the series
\eqn\zetdef{
\zeta(s,\alpha)=\sum_{n=0}^\infty(\alpha+n)^{-s}\ ,
\qquad \Re e\, s>1\ }
to the whole complex plane of the variable $s$. 
This function is analytic  everywhere in the
$s$-plane except the point 
$s=1$, where it has a simple pole. 
Therefore the eigenvalues \psif\ are 
entire functions of $\nu$ in agreement with our conjecture.

The values of \psif\ at the integer
points on the imaginary $\nu$-axis
follow from the formula 
\eqn\zetint{
\zeta(-m,\alpha)=-{B_{m+1}(\alpha)\over m+1}\ ,
\qquad m=0,1,\ldots \ ,}
where $B_m(\alpha)$ are the Bernoulli polynomials \  \Bern.
Note, in particular, that
\eqn\psisim{
\Psi(-i)=1\ ,\qquad \Psi(0)=p\ ,}
in agreement with \triv\ and \psiiq. Further, the values of \psif\ 
at $\nu=i(2n-1)$, $n=0,1\ldots$, conjectured in \psii, 
perfectly  match
the eigenvalues of the local IM, 
which for $\beta^2=\frac{1}{2}$ can be independently
obtained from the explicit expression for the local IM
through free fermion
fields
\eqn\Ifer{
I^{(vac)}_{2n-1}=2^{-n}\ B_{2n}\big(2p+\frac{1}{2}\big)\ .}
The equation \psiir
\ can not be tested since the eigenvalues of the dual
nonlocal 
IM $\widetilde{{\bf H}}_{n}$ are
not generally known independently except for 
the vacuum eigenvalue of
$\widetilde{H}^{(vac)}_1$. For the latter one 
we have
\eqn\psiht{
\Psi^{(vac)}(2i)=2^{-3}\,
B_3\big(2p+\half\big)= p\ \big(p-\frac{1}{4}\big) 
\ \big(p+\frac{1}{4}\big)\ .}
in a precise agreement with \ \psiir\ and
the relation which is  dual  to
\htwoex.

Finally note that the asymptotic expansions of \Tsigma\ and \Asigma\
agree with \ \loexp
\ and \Aas\ provided one calculate the latter in the 
limit $\beta^2=\half-\epsilon$, $\epsilon\to 0$, and identifies the
divergent part of the coefficient in the leading asymptotic of
\loexp\
and \Aas\ with the renormalization constant ${\cal C}$ in \Cval,
\Tvbh\ and
\Avbh.

\newsec{Vacuum eigenvalues of the {\bf Q} - operators and
non-equilibrium states in  boundary sine-Gordon model}

As is explained in Sect.2
the vacuum eigenvalue $A^{(vac)}(\lambda\   )$
for integer $2p = N$ coincides with the Coulomb gas partition
function \ \mjuyt.
In fact it was already observed in\ \SFN 
 \ that the Coulomb gas partition
function \ \mjuyt\
with $p=0$ satisfies the functional equation \ \baxg\foot
{We obtained this result independently,
along with more general statement
\TAp , before the paper \ \SFN\  appeared.
}. The Coulomb gas partition function is
obviously related to the finite temperature
theory of mass-less bose field $\Phi (t,x)$ on
the half plane $x < 0$ with interaction at
the boundary; its action is
\eqn\bsg{{\cal A} = {1\over {4\pi\beta^2}}
\int_{-\infty}^{\infty} dt \int_{-\infty}^{0} dx
\ \big(\,\Phi_{t}^2 - \Phi_{x}^2\,\big)  +
\frac{\kappa}{\beta^2} \  \int_{-\infty}^{\infty} dt
\cos\  \big(\Phi(t,0)+ V t\big)\ .}
Here $\beta^2,  \kappa$
and $V$ are parameters \foot{The
composite field $\cos(\Phi + Vt)$ in \bsg 
\ is assumed to be
canonically normalized with respect to its short-distance
behavior, i.e. 
$\cos\big(\Phi(t)+Vt\big)\cos\big(\Phi(t')+Vt'\big ) \sim
2^{-1}\ \big(\,i (t-t')+
0\, \big)^{-2\beta^2}$,
as is conventional in conformal perturbation theory. This is why no 
ultraviolet cutoff will appear in the matrix elements below.}. 
This model finds interesting applications
in dissipative quantum mechanics\ \leggett,\ \Schmid,\ 
\Fisher,\ \Callan.
As was discussed in\ \Kane,\  \Moon\ 
it also describes the universal current through the
point contact in quantum Hall system.
At nonzero driving potential $V$ and
arbitrary temperature $T$ the system
\ \bsg\  develops a stationary
non-equilibrium state with
\eqn\curr{J_{B} \equiv  {1\over {2\pi}}\langle\, \Phi_x(t,x)
\, \rangle = -  \kappa
\, \langle\, \sin\big(\Phi(t,0)+ V t\big)\, \rangle\  \neq \ 0\ .}
This quantity is interpreted as the backscattering current
through the
point contact \foot{The voltage $V$
and current $J_B$\ 
\curr\  differs 
in normalization from the real voltage
$V^{(phys)}$ and total
current  $J^{(phys)}$\ 
in the Hall system
$$V^{(phys)} = e^{-1}\ V\ ,\ \ \ \ 
J^{(phys)}=\frac{e}{h}\ \beta^2\ \big(\, V+J_B\, \big)\ ,$$
where $e$ and $h$ are the
electron charge  and Plank's constant.
Also, \ $\beta^2$ coincides with the
fractional filling of the Luttinger state
in a   Hall bar 
and the temperature is  measured in  energy units.}.
In general case
\ \curr\  is
non-equilibrium expectation
value and as such it requires non-equilibrium methods
for its computation  \Kane.
Note that if the driving potential is continued to
pure imaginary values
\eqn\vn{V = V_N \equiv 2 \pi i N\  \beta^{-2}  T  }
with integer $N$, one can make the
Wick rotation $t\to - i\tau$ in \ \bsg\
and
formally define the Matsubara partition function
\eqn\part{
Z_{N} = \int[{\cal D}\Phi]\ \exp(-{\cal A}_{M})\ ,}
where
\eqn\mats{{\cal A}_{M} = {1\over {4\pi \beta^2 }}
\int_{0}^{\beta^2/T} d\tau \int_{-\infty}^{0} dx
\ \big(\, \Phi_{\tau}^2 + \Phi_{x}^2\, \big) -
\frac{\kappa}{\beta^2} \ \int^{\beta^2/T}_{0} d\tau
\cos \big(\Phi(0,\tau)+2\pi  N\ \beta^{-2}T\tau \big) }
and the functional integral
is taken over Euclidean fields $\Phi$ which
satisfy
the Matsubara condition
$\Phi(\tau + \beta^2/T,x) = \Phi(\tau,x)$. It is easy
to see that up to overall constant (the partition function
\part\ with $\kappa = 0$) \
\part\  coincides with the series \ \mjuyt\
with
\eqn\mumu{  \mu = {{\kappa}\over{2T}}\ \Big(\,{{\beta^2}\over 
{2\pi T}}\, \Big)^{-\beta^2}\ .}
Moreover, it is possible
to show, using non-equilibrium methods, that
for $V = V_N$ \ \vn\
the expectation values of $e^{\pm i \Phi(t,0)}$ in
\ \curr\  can be calculated in the Matsubara
theory \ \mats\  as
\eqn\exp{ \langle\, e^{\pm i \Phi }
\, \rangle
=\int [{\cal D}\Phi]\
e^{\pm i \Phi (\tau, 0) }\  e^{-{\cal A}_{M}}\Big/ Z_{N}(\mu)\ . }
Of course equilibrium state can not support
nonzero current and indeed it is not
difficult to show that
\eqn\eexp{
\big\{e^{V\tau}  \langle\, e^{i\Phi}\, \rangle
- e^{-V\tau}
\langle\, e^{-i\Phi}\, \rangle
\big\}\big|_{V=V_N} = 0\ .}
Nonetheless it is natural
to expect that the non-equilibrium expectation
value \ \curr\
can be obtained by some kind of analytic continuation of \ \exp\
back to real $V$.
Needless to say this analytic continuation is ambiguous.

In \ \FSN\  the current \ \curr\  is calculated
exactly (for integer values of
$\beta^{-2}$) using the Boltzmann equation
with the distribution
function of charge carriers determined through
Thermodynamic Bethe Ansatz technique.
A conjecture
is proposed 
about exact current for arbitrary
$\beta^2$ and $V$ in the recent paper  \SalF.
It was suggested 
there  to define the ``partition function'' $Z_{2p}(\mu)$ for
$V=4 i\pi p \  \beta^{-2} T $ as certain analytic continuation of
$Z_N$ which uses infinite sum expressions for the coefficients in
\mjuyt\  obtained with the help of Jack polynomials \FSLS\foot{We 
acknowledge a private communication with P. Fendley,
F. Lesage and 
H. Saleur who explained to us how the definition of $Z_{2p}$
in \SalF\  must be understood.}. 
It is conjectured in \ \SalF\ 
that the current \ \curr\  can be expressed
in terms of $Z_{2p}(\mu)$ as
\eqn\their{
J_{B}(V,\, \mu,\,  \beta^2) =i \pi T\    \mu \partial_{\mu}
\log {{Z_{2p} (\mu)}
\over {Z_{-2p}(\mu)}}\ ,\ \ \ \ \ \ \ \
p=-\frac{i\beta^2 V}{4\pi T }\ .}
This conjecture agrees with the earlier conjecture for the linear
conductance in \FSLS. In fact, various checks performed in \SalF\  
suggest that $Z_{2p}$ thus defined satisfies the functional 
relation \TAp.  It is therefore 
very plausible that $Z_{2p}$ coincides with the vacuum
eigenvalue $A_{+}^{(vac)}(\l)$. 

Indeed, we found that the result of \ \FSN\  is in complete 
agreement with the following formula \foot{Again, we have arrived 
at (5.10)  independently,
before the paper \ \SalF\ appeared.
However we were significantly influenced by the conjecture
about the universal
conductance proposed in
\ \FSLS\  and by the results of \FSN .}
\eqn\our{
J_{B}(V,\, \mu,\, \beta^2) =
i\pi T  \  \lambda \partial_{\lambda}
\log {{A_{+}^{(vac)}(\lambda)}
\over {A_{-}^{(vac)}(\lambda)}}\ ,}
where
$$\l=i\ \frac{\sin(\pi \beta^2)}{\pi}\ \mu\ ,\ \ \ 
\ \ \ \ \  p=-\frac{i\beta^2 V}{4\pi T }\  .$$

Namely, the expression for $J_B$ obtained in \FSN\ can be 
written in the form \our\ with $A_{\pm}(\l)$ solving the 
functional equation \TAp. Note that for $2p=N$ both
$A_{+}^{(vac)}$
and $A_{-}^{(vac)}$ coincide with $Z_N$, so each of them
defines certain analytic continuation of $Z_N$ to complex $N$. It is
suggestive to note though that analytic properties of
$A^{(vac)}_{+}$ and $A^{(vac)}_{-}$
as the functions of $p$ are different,
namely $A^{(vac)}_{+}$ is analytic at
$\Re e\  2p > -\beta^2$ whereas $A^{(vac)}_{-}$ is analytic
at $\Re e\
2p < \beta^2$. There is no
clear notion of partition function for a non-equilibrium state, and
therefore it is remarkable that $A^{(vac)}_{+}$
(and $A^{(vac)}_{-}$ as well) admits
interpretation as an ``equilibrium-state''
partition function of the system
similar to\  \bsg\   but with 
additional boundary degree of freedom described
by the ``q-oscillator'' ${\cal E}_{\pm}, {\cal H}$ in \ \kjhg .

According to \ \Aas\  the conjecture \ \our\
implies in particular that the
backscattering current \ \curr\
satisfies the following ``strong-week barrier'' duality relation
\eqn\duali{J_{B}\big(\, V,\mu,\beta^2\,\big)\simeq
- V-  \beta^{-2}\ 
J_{B}\big(\, \beta^2 V,\, C
\mu^{-\frac{1}{\beta^2}},\,  \beta^{-2}\, \big)\ , }
(remarkably, the factor containing
the local IM  in \ \Aas\  cancels in the ratio $A_{+}/A_{-}$)
which generalizes similar relation obtained in \ \FSN\ 
for the case $T=0$
\foot{In fact, exactly this duality relation 
(for the nonlinear mobility
in associated dissipative quantum mechanics problem) was proposed
a while ago in \ \Fisher. The arguments
in \ \Fisher\  are based on ``instanton'' description
of the hopping amplitudes in the strong barrier limit;
from general point
of view this description could be regarded as an approximation.
Therefore it looks
quite remarkable to us that this
relation is indeed exact. 
See also the discussion in \ \FSN.}.
The constant\  $C=C(\beta^2)$\  in\  \duali\  reads explicitly:
\eqn\constant{C(\beta^2)=
\Big(\, \Gamma\big(1+\beta^2\big)\,  \Big)^{\frac{1}{\beta^2}}
\ \Gamma\big(1+\beta^{-2}\big)\ .}

Conventionally, one introduces the renormalized coupling parameter
$X$\ \Kane,\ \FSLS
\eqn\xlambda{X^{2} =2^{1-2\beta^2}\ \sqrt\pi\ \ \frac{
\Gamma(1+\beta^2)}{\Gamma(1/2+\beta^2)}\
\ \mu^2\ ,}
defined in such a way that
\eqn\hig{\partial_{V}J_{B}\big(\, V, X, \beta^2\,\big)\big|_{V=0}=
-X^2+O(X^4)\ , \ \ X^2\to 0\ .} 
According to the formula
\ \duali\ the current
admit the following decompositions:
\eqn\expanx{\eqalign{
J_{B}\big(\, V,\,  X,\,  \beta^2\,\big)& =
-\sum_{n=1}^{\infty} F_n \big(\, V,\,\beta^2\, \big)
\  X^{2n}\cr
&\simeq -V+\beta^{-2}\ \sum_{n=1}^{\infty}
\ F_n \big(\, \beta^2 V,\, \beta^{-2}\, \big)\ 
\ \big(\,  \beta^{-2} K\, \big)^n
\  X^{-\frac{2n}{\beta^2} }\ ,}}
where $F_n $ are certain functions and the constant
$K=K(\beta^2 )$ is  
\eqn\exk{ K(\beta^2 ) = \beta^2\  \bigg({\pi \over 4}\bigg)
^{\frac{1}{2}+\frac{1}{2\beta^2}}
\ \bigg\{\, {{\Gamma^3 \big(1 + \beta^2\big)}\over {\Gamma 
\big(1/2 + \beta^2\big)}}\, \bigg\}^
{\frac{1}{\beta^2}}\ 
  {{\Gamma^3 \big(1+\beta^{-2}\big)}
\over {\Gamma\big(1/2 + \beta^{-2}\big)}}\ .}
Note that this constant determines the leading
$X\to\infty$ 
asymptotic of the  backscattering current,
\eqn\kioh{J_B\simeq 
-V+2  \beta^{-2}T\ K(\beta^2) \
\sinh \frac{V}{2T}\ \frac{\Gamma\big(\,
\beta^{-2}+i\, \frac{V}{2\pi T}\, \big)
\Gamma\big(\, \beta^{-2}-i\, \frac{V}{2\pi T}\, \big)}
{\Gamma^2\big(\, \beta^{-2}\,\big)}\ X^{-\frac{2}{\beta^2}}+
O\big(X^{-\frac{4}{\beta^2}}\big)\ .}

We should stress that the formula \our\ is a conjecture and therefore
any checks would be valuable. As was mentioned above this formula agrees
for integer $\beta^{-2}$ with the TBA solution obtained in
\ \FSN. Although the
calculations in \ \FSN\  are also
based on a conjecture (exact applicability
of the Boltzmann approximation to integrable systems), one could check
\our\  against numerical results obtained from the TBA solution. We have 
done that for $\beta^{-2} =3$ and found an excellent agreement with \kioh.
In particular our numerical value for the constant $K$ is 
\eqn\lskiy{K_{num}(1/3)=3.35485280612...\ ,}
(note that this number is slightly off from the numerical result for 
the same constant
given in \ \FSLSN ) which  is in agreement with the exact value
\eqn\lkmnh{K(1/3)=\frac{3\sqrt2 \pi^{\frac{3}{4}}}{5}\
\biggl(\, \frac{2\, \Gamma(7/6)}{\sqrt3}\, \biggr)^{\frac{15}{2}}=
3.35485280611990...\ .}

Completely rigorous
check can be made in the case $\beta^{-2}=2$,
where the formula \our\
agrees with the explicit calculations in
the free-fermion theory\ \Kane,\ \FSN
\eqn\juy{ J_{B}\big(\, V,\,  X,\,  \frac{1}{2}\,\big)=
-\frac{4  T  X^2}{i\pi} \
\Big(\, \psi\big(\, \frac{1}{2}+\frac{2 X^2}{\pi^2}+
\frac{iV}{4\pi T}\, \big)-
\psi\big(\, \frac{1}{2}+\frac{2 X^2}{\pi^2}
-\frac{iV}{4\pi T}\, \big)
\, \Big)\  ,}
where\ $\psi(x)=\partial_x \log \Gamma(x)$.
Another
interesting limiting case is
the classical limit $\beta^2\to 0 .$
In this limit
the eigenvalues $A_{\pm}^{(vac)}(\l)$ reduce to
(see Appendix B):
\eqn\qclass{A_{\pm}^{(vac)}(\l)\big|_{\beta^2 \to 0}\to 
2^{\pm \frac{\rho}{2}}\ \Gamma(1\pm \rho)\  
X^{\mp \rho}\ J_{\pm\rho}(\sqrt2 X)\ ,}
where the variables
$$ X = \sqrt2\ \beta^{-2}\  \l\ ,\ \ 
\ \ \rho = 2 \beta^{-2}\ p $$
are kept fixed when \ $\beta^2\to 0$.
Substituting this into \our\ one obtains after a little algebra
\eqn\jclass{J_{B}\big(\, V,\,  X,\,  0 \,\big)
= -V+\frac{
2T \sinh\frac{V}{2T}}
{I_{\rho}(\sqrt2 X)\ I_{-\rho}(\sqrt2 X)}\ , \ \ 
\rho=- \frac{iV}{ 2\pi  T} \ ,}
where \ $I_{\rho}(x)$\ is modified Bessel function.
As is known the field theory \bsg\ can be
interpreted in terms of dissipative
quantum mechanics of
a single particle in periodic potential 
\ \leggett,\ \Schmid,\ \Fisher,\ \Callan.
As $\beta^2$
plays the role of
the Planck constant, in the limit $\beta^2 \to 0$ this reduces
to the theory of classical Brownian particle
at finite temperature $T$. In
Appendix C we study associated
Fokker-Planck equation and calculate the
classical current.
This gives additional support to the conjecture \our\ .

As was explained above, the vacuum eigenvalue $A^{(vac)}(\l)$ at $p=0$
coincides with the partition
function of \bsg\  at the temperature $T$ and zero
voltage $V=0$.
Therefore the asymptotic expansion \Aas,\  \Adual\  specialized 
for the vacuum eigenvalue with $p=0$ is essentially
the low temperature 
expansion for the associated free energy.
In particular,
the leading asymptotic of the  heat capacitance
${\cal C}\big(\, X,\beta^2\, \big)$
of the point contact in the quantum Hall
system
at  $T\sim const\,
X^{-\frac{1}{1-\beta^2}}
\to 0$ and zero voltage $V$ reads
\eqn\formula{\eqalign{ &{\cal C}\big(\, X,\beta^2\, \big)\sim
\frac{\Gamma\big(\frac{1}{2-2\beta^2}\big)
\Gamma\big(\frac{2-3\beta^2}{2-2\beta^2}\big)   }
{6\ \sqrt\pi\ \beta^2} \  
\biggl\{\frac{\sqrt\pi\,  \Gamma^3\big(1+\beta^2\big)}{2\beta^4\, 
\Gamma\big(\frac{1}{2}+\beta^2\big)}\biggr\}^{\frac{1}{2-2\beta^2}}\ 
X^{-\frac{1}{1-\beta^2}}\ ,
\ \ \ \ \ \ \ \ \ \ 
0<\beta^2<\frac{2}{3}\ ,\cr
&{\cal C}\big(\, X,\beta^2\, \big)\sim
\frac{(1-\beta^2)^2\ \Gamma^2\big(1+\frac{1}{\beta^2}\big)
\Gamma\big(\frac{3\beta^2-2}{2\beta^2}\big)   }{ \sqrt\pi
\beta^4\ 
\Gamma\big(2-\frac{1}{\beta^2}\big)} \ 
\biggl\{\frac{\sqrt\pi\,  \Gamma^3\big(1+\beta^2\big)}{2\,
\Gamma\big(\frac{1}{2}+\beta^2\big)}\biggr\}^{\frac{1}{\beta^2}}\ 
X^{-\frac{2}{\beta^2}}\ , \ 
\frac{2}{3}<\beta^2  <1\ .}}
Likewise,
\loexp\  determines
the low temperature
expansion for the  impurity free energy 
in the  $s=1/2$ anisotropic  Kondo problem.

\newsec{Discussion}
In this paper we have studied further how the powerful
apparatus of the Yang-Baxter theory of integrability can 
be brought about directly in continuous Quantum Fields Theory.
We have  constructed the operators ${\bf Q}_{\pm}(\lambda)$,
which are field theoretic versions of the $Q$ - matrix of Baxter,
directly in Conformal Field Theory. The ${\bf Q}$ - operators are
constructed as the traces of certain monodromy matrices associated
with the infinite - dimensional spaces - the representation spaces
of ``q-oscillator algebra''\osc. It is worth mentioning that our 
construction is not specific for the continuous theory - the $Q$-
matrix of lattice theory admits similar representation (we will
give the details elsewhere). We also found that the ${\bf Q}$ - 
operators thus constructed
satisfy the remarkable relations \kon,\  \lkjh,\ 
\wron\ and \mnhy. These relations allow one to employ the powerful 
machinery of nonlinear
integral DDV equations  to study the eigenvalues
of the ${\bf Q}$ - operators
in the highest weight Virasoro module. We have 
used the DDV approach
to derive (under some analyticity  conjectures) various
asymptotic expansions
for both ${\bf Q}$ and ${\bf T}$ operators. We also
observed a remarkable (although somewhat puzzling) relation between
the vacuum eigenvalues of the ${\bf Q}$ - operators and 
the stationary transport
characteristics in boundary sine-Gordon model
(the later also relate to the
kinetic properties of one-dimensional quantum particle
coupled to a dissipative environment).
In this paper we did not present the derivation of our
 basic relations \kon,\ \lkjh,
\ \wron\  and \mnhy; this gap will be filled in 
the forthcoming paper \ \BLZn.

Clearly, further study of the
${\bf Q}$ and ${\bf T}$ operators is desirable.
First, almost all the
discussion in this paper concerns the ``Semi-classical
Domain'' $\beta^2< 1/2$
(the case $\beta^2 = 1/2$ is studied explicitly
through the free fermion
theory in the Sect.4). However, the most 
interesting CFT (notably,
the unitary CFT) lay outside this domain. We
have argued in the Sect.2
that the ${\bf Q}$ operators can be defined for
$1>\beta^2 > 1/2$ as well,
but there are reasons to believe that outside
the SD the analytic
properties of both ${\bf Q}$ and ${\bf T}$ operators
undergo significant (and very interesting) changes. We are planning
to extend our analysis to the domain $1>\beta^2 > 1/2$ in the future.

The analysis in this paper is concerned
explicitly with the conformal field
theories (more precisely, the chiral sectors of a CFT). One can argue
however that the operators ${\bf Q}_{\pm}(\lambda)$ (as well as the 
${\bf T}$ operators of \ \BLZ ) can be 
defined in non-conformal integrable QFT obtained by perturbing the
CFT with the local operator $\Psi_{1,3}$, the most important relations
\kon,\  \lkjh,\  \wron\ and \mnhy\  remaining intact, and the most 
significant modifications being in analytic properties of these operators
(they develop essential singularities at $\lambda^2 \to 0$ in the perturbed
case). The ${\bf Q}$ operators and associated (modified) DDV equations
can be used then to study the finite-size spectra of these non-conformal
theories. We will report some preliminary results in this direction in the 
forthcoming paper \ \BLZNN.

And finally it seems extremely desirable to get more understanding about
the relation of the
${\bf Q}$ - operators to the non-equilibrium properties
of the boundary sine-Gordon model discussed in Sect.5 above. In particular
it seems important to find
a physical interpretation to the ``$q$ - oscillator''
degrees of freedom which evidently play central role in our construction
of the ${\bf Q}$ - operators. 

\centerline{}

\centerline{{\bf Acknowledgments}}

V.B. would like to thank R. Baxter for stimulating interest to
this work. 
S.L. is  grateful to  the Department of Theoretical Physics,
Australian National University for the hospitality,
and  would like to thank
A. Belavin, V. Drinfel'd,
J. Jimbo, P. Fendley,  E. Frenkel,
A. LeClair, T. Miwa and   N. Reshetikhin  for interesting discussions.
The work of S.L. is supported in part by NSF grant.
A.Z. is pleased to acknowledge warm hospitality extended to him
in LPM, University of Montpellier and in Laboratoire de Physique
Theorique, ENS, Paris, during
the last stages of this work, and very useful discussions
with Al. Zamolodchikov and V. Kazakov and N. Sourlas.
The research of A.Z.
is supported by Guggenheim Fellowship and by DOE grant 
$\#$DE-FG05-90ER40559\ .

\newsec{Appendix A.} 
In this Appendix we give details of the derivation of the DDV equation
 \ddv \ \dVega.
For any eigenvalue $A(\l)$ satisfying \TAp\ consider the  function $a(\l)$ 
defined by \al.
For 
real values of $p$ the property (ii) of the function $A(\l)$ (given 
in Sect.3 above the formula \sd) implies 
\eqn\reala{
a(\l)^*=a(\l^*)^{-1}\ ,}
where the star denotes the complex conjugation. Next, it follows 
from leading
asymptotic \ytr\ of $A(\l)$ at large $\l$ that  
\eqn\alinf{\log a(\l)
\sim - 2i \, M\  {\rm cos}(\pi\xi/2)\ \l^{\xi+1},\qquad
\l^2\to\infty\ , \qquad
\big|\arg \l^2\big|<2\pi \beta^2\ ,}
where\ $\beta^2=\frac{\xi}{1+\xi}$.
Moreover, the function $a(\l)$ obviously remains finite
for small $\l^2$
\eqn\alsmal{a(\l)=4\pi i p+O(\l^2), \qquad \l^2\to 0\ .}
The function $a(\l)$ satisfies the Bethe-Ansatz type equations \BAe .
The product representation \sd\ implies
\eqn\lhs{\log a(\l)-4\pi ip =\sum_{k=0}^\infty\ F(\l\l^{-1}_k)\ ,}
where 
\eqn\Fdef{\qquad F(\l)=\log{1-\l^2 q^2
\over1-\l^2 q^{-2}}\ .}
The sum in the RHS of \lhs\ can be written as a contour integral
\eqn\integ{\log a(\l)-4\pi i p=
f(\l)
+\int_C{d \mu \over 2\pi i}\ F(\l \mu^{-1})\ 
\partial_{\mu}{\rm log}\big(1+a(\mu)\big)\ ,}
where 
\eqn\fins{f(\l)={\sum_a}' F(\l\l_a^{-1})}
denotes a finite  sum
including only those zeroes $\l_a^2$ which do not lie
on the positive real
axis in the  $\l^2$-plane. 
The contour $C$ goes from $+\infty$ to zero above
the positive real axis, then winds around zero and returns to infinity 
below the positive real axis in the $\l^2$-plane. 
Integrating by parts (boundary terms
vanish due to \alinf, since $0<\xi<1$ for $\beta^2$ in  SD \oiuy) 
one obtains
\eqn\intega{\eqalign{
\integ=&
f(\l)-\int_0^{\infty}{d\mu\over 2\pi i\mu}\ 
\l\partial_{\l}F(\l\mu^{-1})\ 
\bigl\{\log\big(1+a(\mu+i0)\big)-\log\big(1+a(\mu-i0)\big)\bigr\} 
\cr
=\int_0^{\infty}{d\mu\over \pi \mu}&\  \l\partial_{\l}F(\l\mu^{-1})\ 
\Im m \log\big(1+a(\mu-i0)\big)
-\int_0^{\infty}\,{d\mu\over 2\pi i\mu}\
\l\partial_{\l}F(\l\mu^{-1})\ 
\log a(\mu)\ ,}}
where  we have used \reala.

Introducing new variables $\t$, $\t'$ and $\t_k$ as 
\eqn\variab{
\l=e^{\frac{\theta}{1+\xi}},\qquad   \mu=e^{\frac{\theta'}{1+\xi}},
\qquad \l_k=e^{\frac{\theta_k}{1+\xi}}}
and recalling that
$q=e^{\frac{i\pi\xi}{1+\xi }}$,
one obtains
\eqn\integb{
\integ=f(\theta)-
2 i\int_{-\infty}^{\infty}d\theta'\ 
R(\theta-\theta') \  \Im m \log\big(1+a(\theta'-i0)\big) 
+\int_{-\infty}^{\infty}d\theta'\
R(\theta-\theta')
\log\  a(\theta')\ . }
where 
\eqn\defphi{R(\theta)=
\frac{i}{2 \pi (1+\xi)}\ \l\partial_{\l} F(\l)\ .}
With the standard notation for the convolution\ 
\ewq ,
the equation \integb\ can be written as 
\eqn\kleft{
K\ast \log a(\theta)= 4\pi i p +f(\theta)-2 i\  R \ast  \Im m
\log\big(1+a(\theta-i0)\big)\ , }
where
\eqn\defK{K(\theta)=\delta(\theta)-R(\theta)\ .}
Let us now apply the inverse of the
integral operator $K$ to both sides of \kleft; in this one has to add
an  appropriate zero mode of $K$ to the r.h.s. of the resulting
equation  to make it 
consistent with the asymptotic conditions \alinf\ and \alsmal.
In this way one obtains the integral equation
\ddv\ in the main text. As is noted  there, this integral 
equation has to be
complemented by a finite number of the transcendental equations
\BAe\ 
for the roots $\l_a^2$ lying outside the positive real axis of $\l^2$.

For the vacuum eigenvalues with
$2p>-\beta^2$ all the roots $\l_k^2$ are real and positive and the
term $i\sum'_a\log S(\t-\t_a) $ in the r.h.s. of
\ \ddv\ is absent.
Let $\l_0^2$ be the minimal of the zeroes $\l^2_k$. Introduce 
the function $a^{(vac)}(\l)$ related to $A^{(vac)}(\l)$ as in \al\ and
denote 
\eqn\minz{
B(p)={1+\xi\over2}\ \log \l_0^2\ .}
The function $\log\big(1+a^{(vac)}(\t)\big)$ is analytic in the strip
$-\pi\xi <\Im m\ \t<\pi\xi$ 
with the branch cut along the positive real axis from
$B(p)$ to infinity. So it is analytic for real $\theta<B(P)$, where it 
obeys the relation
\eqn\apro{\log a(\theta)=2\   \Im m\, \log\big(1+a(\theta - i0)\big)\ , }
which follows from \reala . Note that the infinitesimal shift $-i0$
here is not essential. Taking \apro\ into account one can rewrite the
\ddv\ for $\t<B(p)$ as 
\eqn\ddvb{\eqalign{-\frac{\pi  p}{\beta^2}
+M\  {\rm cos}\frac{\pi\xi}{2}\  & e^\theta
-\int_{-\infty}^{B(P)}\frac{d\theta'}{2\pi i }\ \partial_{\theta}
{\rm log}S(\theta-\theta') \  
\Im m\, \log\big(1+a^{(vac)}(\theta')\big)\cr=
&\int^{+\infty}_{B(p)}\frac{d\theta'}{2\pi i}\
 \partial_{\theta}
{\rm log}S(\theta-\theta')\
\Im m\, \log\big(1+a^{(vac)}(\theta'-i0)\big)\ .\cr}}
Let us make technical assumption
\eqn\juio{B(p)\sim const\  
{\rm log}\ p \qquad {\rm as} \quad p \to+\infty\ .}
Then one can show that the r.h.s. of \ddvb\ 
decreases at the  large positive  $p$
and therefore can be dropped in the leading 
approximation at $p\to+\infty$. This brings \ddvb\ to the linear
integral equation \lopaa\ of the Winer-Hopf type.

\newsec{Appendix B}
In this appendix we consider the functional
equations  for the eigenvalues of ${\bf T}$ and ${\bf Q}$ operators
in the classical limit $\beta^2\to0$, where they reveal a remarkable 
connection with the theory of classical
Liouville equation \foot{
The fact that the functional relations  
associated with the TBA equations
have many features in
common with the classical Liouville equation is well known to 
experts \ \Zamt. }. We start with the functional equations for
the eigenvalues $T_j (\l)$ of the operators \nhy\ which follow
from \fus\ ; they can be written in the form \BLZ
\eqn\cdsx{T_{j}(q^{\frac{1}{2}}\l)T_{j}(q^{-\frac{1}{2}}\l)=1+
T_{j-\frac{1}{2}}(\l)T_{j+\frac{1}{2}}(\l) } 

Consider limiting values of the
eigenvalues  $T_j(\l)$ and $Q_\pm(\l)$,
when
\eqn\clvar{
2\l=\beta^2\  e^\sigma, \qquad j={\tau\over\pi\beta^2}, \qquad
2p=\beta^2\ \rho,\qquad \beta^2\to  0\ ,}
and the variables $\sigma$, $\tau$ and $\rho$ are kept fixed. We assume
that in this limit
\eqn\tqlim{T_j(\l)\to {2\over\pi\beta^2}\ 
e^{-\phi(\sigma,\tau)},\qquad Q_\pm(\l)\to
\big(\, {\beta^2}\, \big)^{\pm \rho}\  \cQ_\pm(\sigma)\ ,}
where $\phi(\sigma,\tau)$ and $\cQ_\pm(\sigma)$ are smooth functions
of their arguments. Then it is easy to see that the functional
equation \cdsx\ becomes the classical Liuoville equation
\eqn\Liou{
(\partial^2_{\sigma}+\partial^2_{\tau})\
\phi(\sigma,\tau)=e^{2\phi(\sigma,\tau)}\ }
for the field $\phi(\sigma,\tau)$ in the Euclidean space.
It should be
complemented by the periodical boundary conditions
\eqn\fper{\phi(\sigma+i \pi ,\tau)=\phi(\sigma,\tau)\ ,}
since $T_j(\l)$ in \tqlim\ is a single-valued function of $\l^2$.
The limiting form of the functional equation \mnhy\ reads
\eqn\Lsol{e^{-\phi(\sigma,\tau)}=(4 i \rho )^{-1}\ \big(\
\cQ_+(\sigma+i \tau)\cQ_-(\sigma-i \tau)-
\cQ_+(\sigma-i \tau)\cQ_-(\sigma+ i \tau)\ \big)\ ,  }
whereas \wron\ becomes the ordinary Wronskian condition
\eqn\cwron{\partial_\sigma\cQ_+(\sigma)\ \cQ_-(\sigma)-
\cQ_+(\sigma)\ \partial_{\sigma}\cQ_-(\sigma)= 2 \rho\ .}
for the function\ $ \cQ_{\pm}(\sigma)$.

One can  recognize in\ \Lsol,
\cwron\   the general local solution of the
Liouville equation   \Liou\ 
satisfying the condition
\eqn\condi{
e^{-\phi(\sigma,\tau)}=\tau+O(\tau^2)\ ,\qquad \tau\to 0.}

The Baxter's relation \baxg\ reduces to the second order 
differential equation
\eqn\diffe{\big(\partial^2_{\sigma}+w(\sigma)\big)\ \cQ(\sigma)=0\ ,}
with a periodic potential
\eqn\pper{w(\sigma+i\pi )=w(\sigma)\ ,}
which is determined by the leading asymptotic of
the eigenvalue $T(\l)\equiv T_\half(\l)$ in the limit \clvar\
\eqn\wdef{T(\lambda)= 2+(\pi \beta)^2\ w(\sigma)\ +O(\beta^4)}
The functions\  $\cQ_{\pm}(u)$ are just two linearly
independent Bloch-wave  solutions
\eqn\loc{\cQ_{\pm}(\sigma+i \pi )=
e^{\pm \pi i  \rho}\  \cQ_{\pm}(\sigma)\ .}
to the second order
differential equation \diffe.
Note that  \loc\ imply the periodicity \fper\
of the solutions \Lsol\ for the Liuoville equation \Liou.

It is illustrative 
to apply the above limiting procedure to the vacuum
eigenvalues of ${\bf T}$ and ${\bf Q}$ operators.
Notice  that all the vacuum
eigenvalues of the nonlocal IM \mjui\ entering the series expansion
\knhdy\ remain finite in the limit \clvar.
Therefore only two first terms of
the expansion
\eqn\Texp{T^{(vac)}(\l)=
2\cos(2\pi p) + \l^2 G_1^{(vac)} + O(\l^4)\ .}
contribute to \wdef. Using \qvac\ one thus obtains
\eqn\wvac{w(\sigma)=e^{2 \sigma}-\rho^2\ }
The solutions to the differential equation
$$ (\partial^2_{\sigma}+e^{2 \sigma}-
\rho^2)\ \cQ_{\pm}(\sigma)=0\ ,$$
satisfying  the conditions\
\cwron, \loc\  have the   forms
$$\cQ_{\pm}(\rho)=
\Gamma(1\pm \rho)\  J_{\pm \rho}(e^\sigma)
\ ,$$
where\ $J_\sigma(x)$\ is the conventional  Bessel function.

The eigenvalues $ T^{(vac)}_j(\l)$ in the limit \clvar\
are then expressed from \tqlim\ through the corresponding solution
\Lsol\ of the Liouville equation
\eqn\Tcl{\eqalign{ T^{(vac)}_j(\l)\to&\ 
\frac{2}{\pi \beta^2}\ e^{-\phi(\sigma,\tau)}=\cr
&\frac{1}{2i\beta^2 \ {\rm sin}\pi \rho} \
\big(\ J_\rho(e^{\sigma+i\tau})
J_{-\rho}(e^{\sigma-i\tau})-J_\rho(e^{\sigma-i\tau})
J_{-\rho}(e^{\sigma+i\tau})\ \big)\ .}}
It would be interesting to check this formula against the solutions
to the TBA equations in the limit $\beta^2 \to 0$.

\newsec{Appendix C}

As is discussed in  \leggett\  the field theory \bsg\ 
describes a dissipative 
quantum mechanics of one dimensional particle if one interprets
the boundary value $\Phi(t) \equiv \Phi(t,0)$ as the position of
the particle, the bulk degrees of freedom $\Phi(t,x), \ x \neq 0$
playing the role of thermostat. As $\beta^2$ enters \bsg\  as the
Planck's constant, in the limit $\beta^2 \to 0$ this problem reduces
to the one of classical Brownian particle which is described by the
Langevin equation
\eqn\langevin{{1\over {2\pi}}
\ {\dot\Phi(t)} = -\kappa\ \sin\big(\Phi(t) + Vt\big) + \xi(t),}
where ${\dot\Phi} = \partial_t\Phi(t)$,
and $\xi(t)$ is the white noise
\eqn\white{\langle\, \xi(t) \xi(t')\, \rangle =
{T\over \pi}\   \delta(t-t')\ .}
We are interested in the limiting value of the average velocity
\eqn\clcurr{J_B(V)=
\langle\, { \dot \Phi }\,\rangle_{t\to\infty}\ .}
The time dependence of the potential term in \langevin\ can be
eliminated by the substitution
\eqn\phix{z(t) = \Phi(t) + Vt\ ,}
which transforms \langevin\ to the form
\eqn\langev{ {\dot z} = V - 2\pi \kappa\ \sin(z) + 2\pi\xi\ .}
The driving force $V$ appears here because the transformation \phix\ 
brings us to the frame which moves with the velocity $-V$ with respect
to the thermostat. The current \clcurr\ then is
\eqn\cltok{J_B(V) = -V + J (V)\ , \qquad J(V) = 
\langle\, {\dot z}\, \rangle_{t\to\infty}\ .}
The probability distribution $P(z)$ 
associated with the stochastic process
\langev\ satisfies the Fokker-Planck equation
\eqn\fokker{ \partial_t P =
2\pi T\  \partial_z \big\{\,(-\nu + \sqrt2 X\ \sin z)\ P + 
\partial_z P\, \big\}\ ,}
where
\eqn\nuz{\nu = {V\over{2\pi T}}\ ,
\qquad \sqrt2 X = {\kappa\over T}\ .}
The suitable solution of \fokker, 
which describes the stationary drift has
the form (see e.g.  \Stratonovich)
\eqn\prbab{P(z) = {\cal N}^{-1}\  P_0 (z)\ , \ \ 
P_0 (z) = e^{\sqrt2 X\cos(z) + \nu z}
\int_{z}^{z+2\pi} {{dy}\over{2\pi}}\ 
 e^{-\sqrt2 X \cos(y) - \nu y}\ ,}
where 
\eqn\gtre{{\cal N}= \int_{0}^{2\pi} dz\  P_0 (z)  }
is the normalization constant.
The current $J$ can be
then expressed as
\eqn\jnol{J =
2\pi T\  {\cal N}^{-1}\ \big(1-e^{-2\pi\nu}\big)\ .}
The integral\ \gtre\ 
is calculated explicitly with the result
\eqn\jnolf{J =
{ 2T\  {\sinh(\, \pi\nu\, )}\over {I_{i\nu}(\sqrt2 X)\ 
I_{-i\nu}(\sqrt2 X)}}\ ,}
where $I_{\rho}(x)$ is the modified Bessel function.
With the account \ \cltok,\ \nuz,\
\mumu,\  \xlambda,    this agrees perfectly with \jclass.

\listrefs

\end